\documentclass[11pt]{article}
\usepackage{geometry}                
\usepackage[parfill]{parskip}    
\usepackage{graphicx}
\usepackage{amsmath, amsthm, amsfonts}
\usepackage{amssymb}

\newcommand{\sect}[1]{\setcounter{equation}{0}\section{#1}}

\usepackage{color}
     \usepackage[colorlinks]{hyperref}
\hypersetup{linkcolor=blue,%
citecolor=red,%
urlcolor=cyan}
\usepackage{url} 

\usepackage{epstopdf}
\begin{document}

\title{Electric and magnetic waveguides in graphene:
\\ 
quantum and  classical}

\author{
David Barranco$^b$\footnote{davidbarran95@gmail.com, ORCID: \href{http://orcid.org/0009-0001-7405-1543}
{0009-0001-7405-1543}}, 
\c Seng\"ul Kuru$^a$\footnote{sengul.kuru@science.ankara.edu.tr, ORCID: \href{http://orcid.org/0000-0001-6380-280X}{0000-0001-6380-280X}}, 
Javier Negro$^b$\footnote{jnegro@uva.es, 
ORCID: \href{http://orcid.org/0000-0002-0847-6420}{0000-0002-0847-6420}}
\\ 
$^a$Department of Physics, Faculty of Science, Ankara University, 06100 Ankara, T\"urkiye\\
$^b$Departamento de F\'i{}sica Te\'orica, At\'omica y Optica, Universidad de Valladolid,\\ 47011 
Valladolid, Spain\\}

	\maketitle
	
\begin{abstract}

Electric and magnetic waveguides  are considered in 
planar Dirac materials like graphene as well as their classical version for relativistic particles 
of zero mass and electric charge.
In order to deal with analytical solutions of the Dirac-Weyl equation, we have assumed 
the displacement symmetry of the 
system along a direction. Afterwards, we have examined the rest of symmetries 
relevant to each type of waveguide, magnetic or electric, which will determine their main  differences. We have worked out waveguides 
with square profile in detail to show up some of the 
most interesting features. This is made much more clear in the light of a background classical analog. By means of a systematic comparison of both types of waveguides it has been possible to elucidate their more relevant aspects. Most  results have been visualized along a series of representative graphics.
\end{abstract}

\sect{Introduction}
Graphene and 2D Dirac materials have been a centre of attention beyond pure condensed matter researchers in the last years due to some of its peculiar electronic and optical properties in experimental and theoretical physics \cite{Neto10}. In particular, there has been an considerable number of works concerned about the effect of external electric and magnetic fields on  transport and confining properties of Dirac electrons. However, electric and magnetic fields, give rise to quite different influences on Dirac electrons. One of the key points is the strong Klein tunnelling effect \cite{Katsnelson06,superKlein20} which is most relevant in electric interactions. This leads to difficulties to achieve confinement under electric fields although it can be reached under some special conditions \cite{vit22}; on the other hand, although magnetic fields seems more appropriate on this aspect, it happens that too intense magnetic fields may also have undesirable effects in this respect.

Many contributions have been devoted to the confining problem in bounded regions, that is in quantum dots, and in one dimensional regions like quantum waveguides, as well as to their  scattering properties on a wide variety of field arrangements. This work is concerned with a detailed discussion about the differences that electric and magnetic waveguides may present.
In spite of the numerous previous papers on  waveguides, we thought it convenient to give a basic and global description about the fundamental and characteristic  aspects of electric and magnetic fields in the waveguide configuration which should be taken into account.

We will start with a classical approach which is not usual, but it will prove to be quite clarifying afterwards, in the quantum frame.
We have taken three characteristic parameters determining both quantum and classical systems of massless particles: the energy $E$, the momentum $k$  in the $y$ direction (symmetry direction) and the depth  or intensity of the potential ($v_0$ for electric  or $a_0$ for magnetic) choosing apapropriate units. We have discussed the influence of these parameters on the scattering and bounded motion of particles for both, scalar (electric) and vector (magnetic) potentials in the classical setup. We have taken in this work  simple rectangular profile shapes (mainly wells) which allow us to obtain immediate results. For example, we have determined the regions of these parameters allowing for bounded  or scattering motion, which are essentially different for the electric and magnetic cases. Some diagrams show what kind  of states have particles with different values of parameters $E$, $k$ and $a_0$ or $v_0$. The classical particles behave similarly to light when it passes from one region to another one with different refraction index by the Snell's law, as has been already mentioned in the literature. What we show here is the details of the different refraction properties of electrons for electric or magnetics waveguides, in particular the asymmetric behaviour in  the magnetic case which depends on the field intensity.

Next, once introduced the classical viewpoint, we  investigated these questions in the quantum context by means of the two-dimensional Dirac-Weyl equation for both electric and magnetic interactions paying attention to scattering and bound states. In this program, the previous considerations on the classical  behaviour constitute a good reference to the quantum problem. For example; i) In the quantum case bound states are described by spectral curves, which in general are inside  the continuous region obtained in the classical study (they become a discrete quantization). ii) These spectral curves in the quantum problem for the vector potential have some especial bound states called edge states which are out of the classical region for this problem (they belong to a classically forbidden region). The edge states are missing in electric waveguides. iii) These two problems (scattering and bound states),  have also been connected by a join plot where spectral curves of the discrete spectrum are represented together with resonances of scattering states ($T=1$) for electric and magnetic potentials. These graphics give some suggestions about the capture of bound states ``atomic collapses'' into the hole sea. This is quite different in the electric and magnetic contexts.

We remark that this work contains many of the results that can be found scattered in the literature. However, we have made a unified presentation which we hope it will allow to appreciate the constrast of these two systems presented under the same approach. We have included many details that can only be fully displayed within our global program. We hope that this work will raise some interest in open questions which are present along our development. 

We wish to mention just a few representative papers close to this work which, in some way, have influenced our presentation. 
Part of the first research which were aware of the importance of magnetic fields in graphene are  the Refs \cite{deMartino07,deMartino07b,deMartino09}, see also \cite{Goerbig11}. An extensive study of different configurations with magnetic fields can be found in refs.~\cite{peeters09,peeters10,Moldovan18}. An interesting paper  \cite{gosh09} is very close to our point of view.
Solutions of different inhomogeneous magnetic fields were obtained by SUSY methods in \cite{kuru09} (see also \cite{Milpas11}). Some cases of electric waveguides are worked out in \cite{peeters06,peeters10b,portnoi14,portnoi17}. The join action of electric and magnetic fields were investigated in a series of contributions \cite{lukose07,Ghosh19,Tan10,roy20,Bautista20,Do21,Afshari17} and in particular a precedent paper \cite{kuru22} to this work.

\sect{Interactions with symmetry in the y-axis direction}
	
We will restrict here to the stationary Dirac-Weyl equation in the plane for quasi-particles of zero mass and $1/2$-spin with the electron charge. This equation corresponds to the continuous approximation to describe the electronic interaction in graphene for a range of energies near  the Dirac points (we will assume the interaction will not mix both, so that hereafter we will restrict to one of them, the point  $K$)
\cite{Neto10}. The equation for the interaction in external static fields, applying the minimal coupling rule, is
\begin{equation}\label{dirac1}
v_F\big(  {\boldsymbol \sigma \cdot}  ({\bf p}+e{\bf \tilde{A}}) \big)\Psi({\bf x}) = \big(\epsilon - \tilde{V}({\bf x})\big) \Psi({\bf x})
\end{equation}
where, as usual ${\boldsymbol \sigma} = (\sigma_x,\sigma_y)$ are sigma Pauli matrices,
${\bf p}=-i\hbar\bf{\nabla}$ the two-component momentum operator in the $xy$-plane, $v_F$ the Fermi velocity of graphene, $e$ the electric charge and ${\bf \tilde{A}}({\bf x})$, $\tilde{V}({\bf x})$  magnetic and electric potentials defined on the plane; the wavefunction
$\Psi({\bf x})=\Psi(x,y)$ is a two-component pseudospinor and $\epsilon$ corresponds to energy. 

We will consider that the fields have a symmetry under displacements along one direction (in the $y$-direction) and both fields to be perpendicular (the electric on the $xy$-plane, the magnetic in the $z$-direction normal to the graphene plane). In fact, we will restrict to  potentials having the form
\begin{equation}
{\bf \tilde{A}}({\bf x}) = (0,\tilde{A}(x)),\qquad \tilde{V}({\bf x})=\tilde{V}(x)
\end{equation}
Therefore, the magnetic field may depend only on $x$ in this case and it is given by 
\[
{\bf \tilde{B}}(x)= \tilde{A}'(x) \hat z=\frac{d\tilde{A}(x)}{dx}\hat z
\]
where $\hat z$ is the unit vector of $z$-axis;
while the electric field 
\[
{\bf\tilde{E}}(x)= -\tilde{V}'(x) \hat x=-\frac{d\tilde{V}(x)}{dx}\hat x
\] 
will be in the $x$-direction ($\hat x$ is the unit $x$-vector) and it may also depend only  on $x$. Hereafter, 
the ``prime", will be used for derivatives with respect to the argument.

Let us make the following definitions in order to simplify the notation and determine the units for $A(x), V(x)$ and $\epsilon$:
\begin{equation}\label{def}
A(x)=-\frac{e \tilde{A}(x)}{\hbar}, \qquad {V}(x)= \frac{\tilde{V}(x)}{\hbar v_F}=-\frac{e\varphi(x)}{\hbar v_F},\qquad E=\frac{\epsilon}{\hbar v_F}
\end{equation}
Thus, equation (\ref{dirac1}) can be rewritten as
\begin{equation}\label{dirac2}
H({\bf x})\Psi({\bf x})=\big(\sigma_x(-i \partial_x)+\sigma_y(-i \partial_y-A(x)) +V(x) I \big)\Psi({\bf x}) = E \Psi({\bf x})
\end{equation}
where $I$ is for the unit matrix. This is the form that we will apply in waveguide configurations. $H({\bf x})$ is the Dirac-Weyl Hamiltonian under the external magnetic $A(x)$ and electric $V(x)$ potentials, but in this work we will discuss each field separately. 

In the following we will list some symmetries of $H({\bf x})$ depending on the parameters $E$, $k$ and $v_0/a_0$ which enjoy the  potentials of our waveguides. They will greatly facilitate the finding and characterization of solutions.

\noindent
\subsection{Symmetries}

{\it  For both interactions}

\begin{itemize}
\item[(i)] {\it  Translation in the $y$-direction}

As we mentioned in the precedent section, the potentials $A(x)$ and $V(x)$ will depend only on $x$, so we can find solutions of the stationary equation (\ref{dirac2}) that at the same time are eigenfunctions of the translation generator in the $y$-direction $P_y= -i\partial_y$. If the eigenvalue of $P_y$ is $k$ then, these eigenfunctions will have the form
\begin{equation}
\Psi(x,y) = e^{i  k\, y}\,\psi(x)= e^{i k\, y}\,\left(
\begin{array}{c}\psi_1(x)
\\[0.2ex]
\psi_2(x)\end{array}\right)
\end{equation}
where $\psi_1(x), \psi_2(x)$ are components of the reduced spinor dependent only on $x$.

\item[(ii)] {\it Reflection in $x$}

We will consider symmetric  potential (for both electric and magnetic) wells: $A(x) = A(-x)$ and
$V(x) = V(-x)$. In such conditions the stationary Dirac equation (\ref{dirac2}) implements this symmetry by means of the operator
\begin{equation}\label{reflection}
{\cal R}_x= \sigma_y\,R_x 
\end{equation}
where $R_x\psi(x) = \psi(-x)$ is the $x$-reflection operator and $\sigma_y$ is the corresponding Pauli matrix. This means that if $\Psi$ is a solution of eq.~(\ref{dirac2}), then ${\cal R}_x\Psi$ will be another solution (may be the same if the symmetry is not broken):
\[
H({\bf x})\Psi({\bf x})=E\Psi({\bf x}) \ \implies\ 
H({\bf x})\big(\sigma_y\,R_x\Psi({\bf x})\big)=E\big(\sigma_y\,R_x\Psi({\bf x})\big)
\]
\end{itemize}

{\it For pure magnetic interactions} 

\begin{itemize}
\item[(iii)] {\it 
(a) Symmetry of positive and negative spectrum}

If $\Psi(x,y)$ is a solution with energy $E$, then $\sigma_x\,R_x \Psi(x,y)$ will be another solution with energy $-E$. Therefore, the spectrum in the magnetic interaction will be symmetric in positive and negative energies:
\[
H({\bf x})\Psi({\bf x})=E\Psi({\bf x}) \ \implies\ 
H({\bf x})\,\big(\sigma_x R_x\Psi({\bf x})\big)=-E\,\big(\sigma_x R_x\Psi({\bf x})\big)
\]

{\it 
(b) Symmetry under $k\to -k$ and $A(x) \to -A(x)$}

If $\Psi(x,y)$ is a solution with energy $E$, momentum $k$ and potential $A(x)$, 
then $\sigma_x  \Psi(x,y)$ will be another solution with the same energy $E$ for
 momentum $-k$ and potential $-A(x)$ :
\[
H({\bf x}, k, A(x))\Psi({\bf x})=E\Psi({\bf x}) \ \implies\ 
H({\bf x}, -k, -A(x)))\,\big(\sigma_x  \Psi({\bf x})\big)=E\,\big(\sigma_x \Psi({\bf x})\big)
\]

\end{itemize}

{\it For pure electric interactions} 

\begin{itemize}

\item[(iv)] {\it 
(a) Symmetric spectrum of $V(x)$ and $-V(x)$}

If $\Psi(x,y)$ is a solution with energy $E$, for the electric potential $V(x)$ then $\sigma_z \Psi(x,y)$ will be a solution with energy $-E$ for the potential $-V(x)$. Therefore, the spectrum of the potential well $V(x)$ and the opposite barrier $-V(x)$, will be symmetric:
\[
H({\bf x}, V(x))\Psi({\bf x})=E\Psi({\bf x}) \ \implies\ 
H({\bf x}, -V(x))\,\big(\sigma_z  \Psi({\bf x})\big)=-E\,\big(\sigma_z  \Psi({\bf x})\big)
\]

{\it 
(b)  Symmetric spectrum for $k$ and $-k$}

If we reverse the sign of the conserved momentum $k\ \to \ -k$, then the eigenfunctions 
transformed under $\sigma_y$ will have the same eigenvalue:
\[
H({\bf x},k)\Psi({\bf x})=E\Psi({\bf x}) \ \implies\ 
H({\bf x},-k)\,\big(\sigma_y \Psi({\bf x})\big)= E\,\big(\sigma_y \Psi({\bf x})\big)
\]

\end{itemize}

\sect{Classical waveguides}

In this section, we will deal with classical waveguides  defined on a plane, which are produced by pure electric or magnetic potentials of square shape for zero mass particles with the electron charge. This is a limiting case of ultrarelativistic charges.

In all the cases, as mentioned earlier, we will have three constant  characteristic values: (i) energy $E$, (ii) momentum in the symmetry direction $k$, and (iii) depth or intensity of the potential ($v_0$ for electric and $a_0$ for magnetic potentials). 
Although in the classical relativistic frame the energy always is positive, we will extend it to the negative sign for some situations explained below.

\noindent
\subsection{The case of an electric well}

Consider the case of an electric waveguide with potential of square profile given by 
\begin{equation}\label{elect}
{V}(x,y)= 
\left\{\begin{array}{cl}
- v_0,  \ &|x|<1 \,, 
\\[1.5ex]
 0,  \ &|x|>1
 \end{array}\right.
\end{equation}
The width of the well has been fixed to simplify the discussion and fix units. We  will always assume that $v_0>0$, so the above potential will be a well (if $v_0<0$, this becomes a barrier. From the symmetries above mentioned, this situation is equivalent to a barrier, changing the sign of energy of the well).
The potential determine three regions in the $x$ variable: $I$ for $x<-1$, $II$ for $-1<x<1$ and $III$ for $x>1$.


Let a classical relativistic charged particle with zero mass be incident from the left towards the well. Then, we have the following possible situations. Firstly we describe the situation of scattering states where a particle incident from the left (region I) into the well will cross it (through region II) to pass to region III.

\noindent
\subsubsection{Scattering through an electric well} 

\begin{figure}[h!]\label{energy1}
\begin{center}
\includegraphics[scale=0.35]{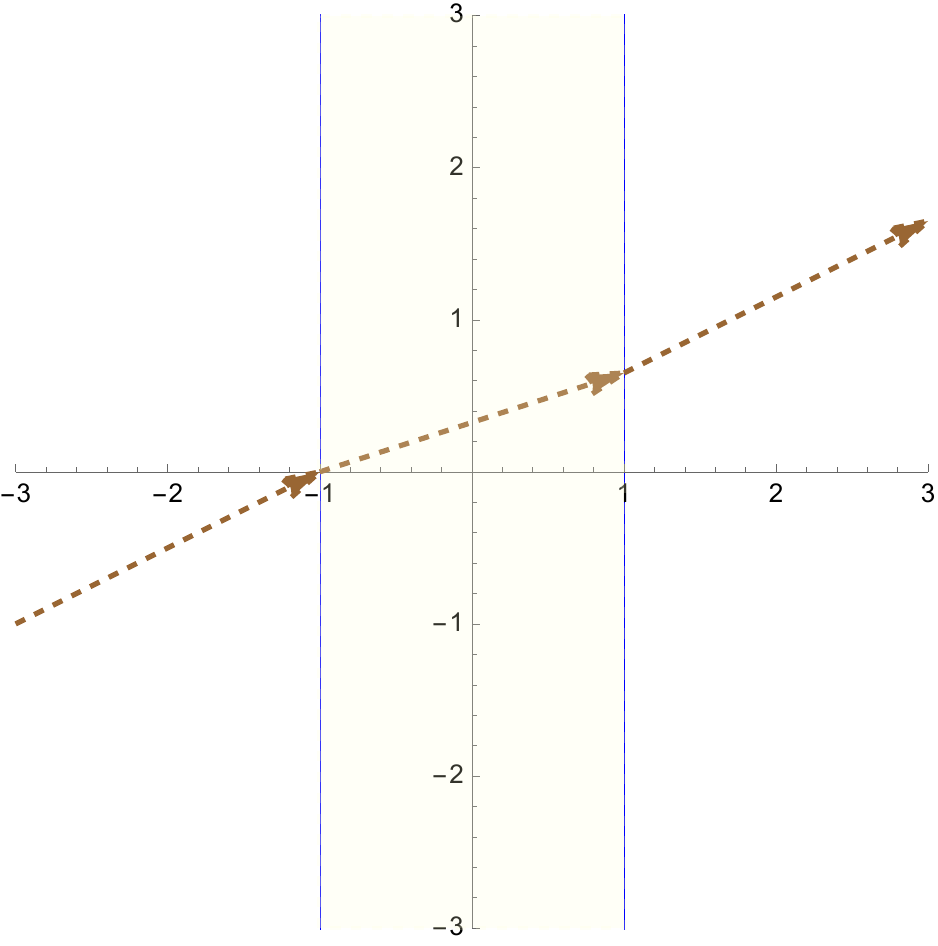}
\includegraphics[scale=0.35]{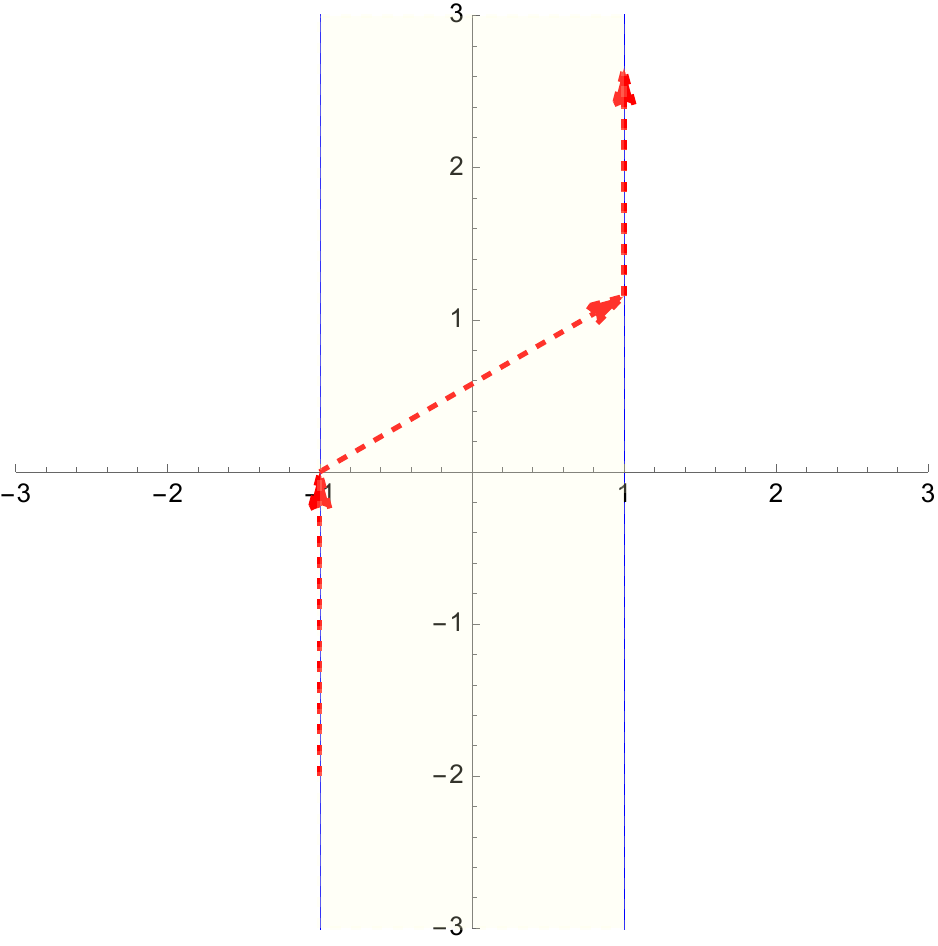}
\includegraphics[scale=0.35]{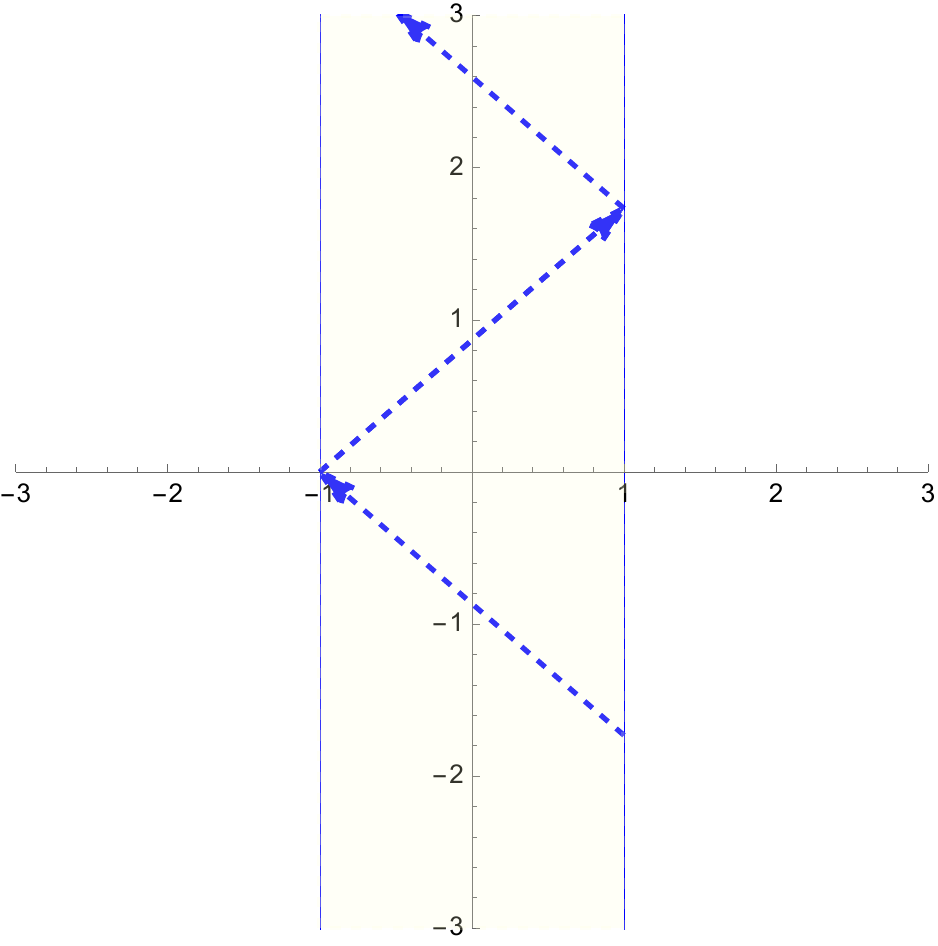}
\caption{\small Trajectories of a particle in an electric well for
the configuration $k=1$, $v_0=1$ (a) $\sin \alpha'< 1/2$, (b)  $\sin \alpha'_{\rm max}= 1/2$  and (c) $\sin \alpha'> 1/2$ for a trapped motion.  \label{figA}}
\end{center}
\end{figure}

\begin{itemize}

\item{\it Regions I and III. $(p_x>0)$}

The incident particle in region I (and in the outgoing region III) has energy $E$ and momenta $p_x$ and $p_y$. Since in the $y$-direction the potential is constant, we have a translation symmetry in that direction, the momentum $p_y$ will be a constant of the motion which we  call $k\equiv p_y$. The potential in the left and right regions is zero, so we have the relativistic zero mass relation
(we take $c=1$)
\begin{equation}\label{r10}
E^2= p_x^2 + k^2 \ \implies p_x=\sqrt{E^2-k^2}
\end{equation}

We will suppose that $p_x>0$; in principle we also assume that $k>0$ and $E>0$, but we may as well include, without additional complications, negative values of $k$ and $E$. So that once $p_y=k$ is fixed, then from (\ref{r10}) 
\[
|E| \geq |k|\,,\qquad 0\leq p_x < \infty
\]

\item {\it Region II. $(p'_x> p_x)$}

Once the particle passes to region II, the effective energy $E$ changes due to the potential of the well: $E'=E+v_0$, while the momenta will be $p'_x$ and $p'_y=k$ (which has the same constant value because the symmetry in the $y$ direction persists).  The energy-momenta relation in region II is now
\begin{equation}\label{r2}
(E+v_0)^2= (p'_x)^2 + k^2 \ \implies p'_x=\sqrt{(E+v_0)^2-k^2}\ > p_x
\end{equation}
If $E$ is positive, this means that inside II the $x$--momentum has increased: $p'_x> p_x$ (Fig.~\ref{figA}, left).
Recall that the initial energy $E$ is greater than $k$ because the particle comes from region I where $p_y=k$.
Then the minimum value of $p'_x$  is reached if $E=k$ (see Fig.~\ref{figA}, center) and therefore in that case, substituting in (\ref{r2}), 
\[
p_x=0 \ \implies\ p'_x=p'_{\rm min} = \sqrt{(k+v_0)^2-k^2}
\]

In conclusion, there will be classical scattering under an electric well for any incident particle characterized by the vertical momentum $k$ and energy $E^2> k^2$.
\end{itemize}
\medskip

\noindent
\subsubsection{Incident and ``refracted'' directions}

We may introduce the angles $\alpha$ for the incident particle in region I and $\alpha'$ for the particle entering in the well region II. In this subsection we are assuming that $E$ is positive. 
These angles $\alpha$ and   $\alpha'$ are defined by 
\begin{equation}\label{r2b}
\begin{array}{ll}
\displaystyle
\sin \alpha = \frac{k}{E}\, ,\qquad 
&
\displaystyle\sin \alpha' = \frac{k}{E+v_0}
\\[2.ex]
\displaystyle
\tan \alpha =  \frac{k}{p_x}=\frac{k}{\sqrt{E^2-k^2}}\, ,
\quad 
& 
\displaystyle\tan \alpha'=  \frac{k}{p'_x}=\frac{k}{\sqrt{(E+v_0)^2-k^2}} , \ \implies \ \alpha>\alpha'
\end{array}
\end{equation}
Therefore,
\begin{equation}\label{r3}
p_x\,\tan \alpha = p_x' \tan \alpha'\,,\qquad (\alpha\geq \alpha')
\end{equation}
or
\begin{equation}\label{r3}
E\,\sin \alpha = (E+v_0) \sin \alpha'\,,\qquad (E\geq k)
\end{equation}
This is similar to Snell's law $n_1\sin \theta_1 = n_2 \sin \theta_2$ of refractive indexes and angles for light waves.
In particular for the minimum energy  $E= k$, $\alpha=\pi/2$, there is a maximum angle $\alpha'_{\rm max}$ entering into the well: 
\begin{equation}\label{r4}
\sin \alpha'_{\rm max} = \frac{k}{k+v_0}
\end{equation}

Thus, we have the following ranges of  angles $\alpha$ and $\alpha'$:
\begin{equation}\label{rangea}
0\leq \alpha \leq \pi/2\,,\qquad 0\leq \alpha' \leq \alpha'_{\rm max}
\end{equation}

The possible trajectories of a classical particle in an electric well are plotted
in Fig.~\ref{figA}, for the three cases: (left) scattering, (center) limiting case ($\alpha=\pi/2$, $\alpha'= \alpha_{\rm max}'$),
(right) a bounded motion.


\noindent
\subsubsection{Bounded motion inside the electric well} 

\begin{figure}[h!]\label{energy2}
\begin{center}
\includegraphics[scale=0.35]{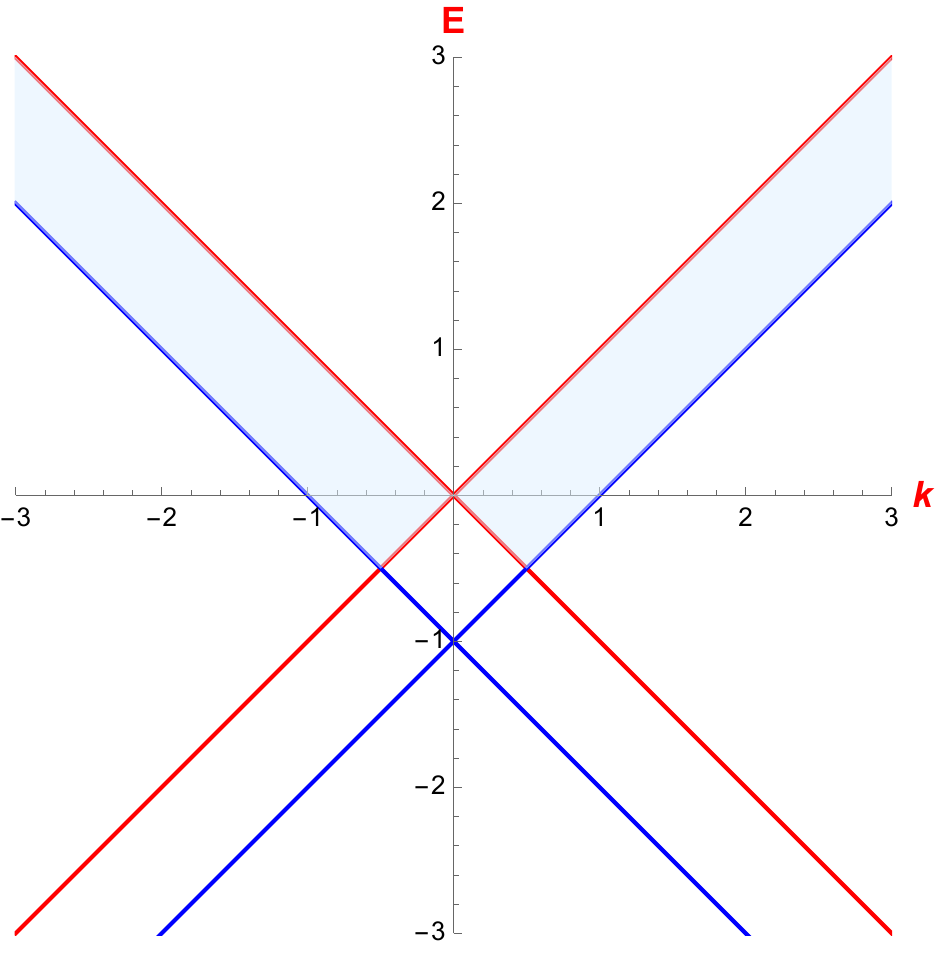}
\qquad
\includegraphics[scale=0.35]{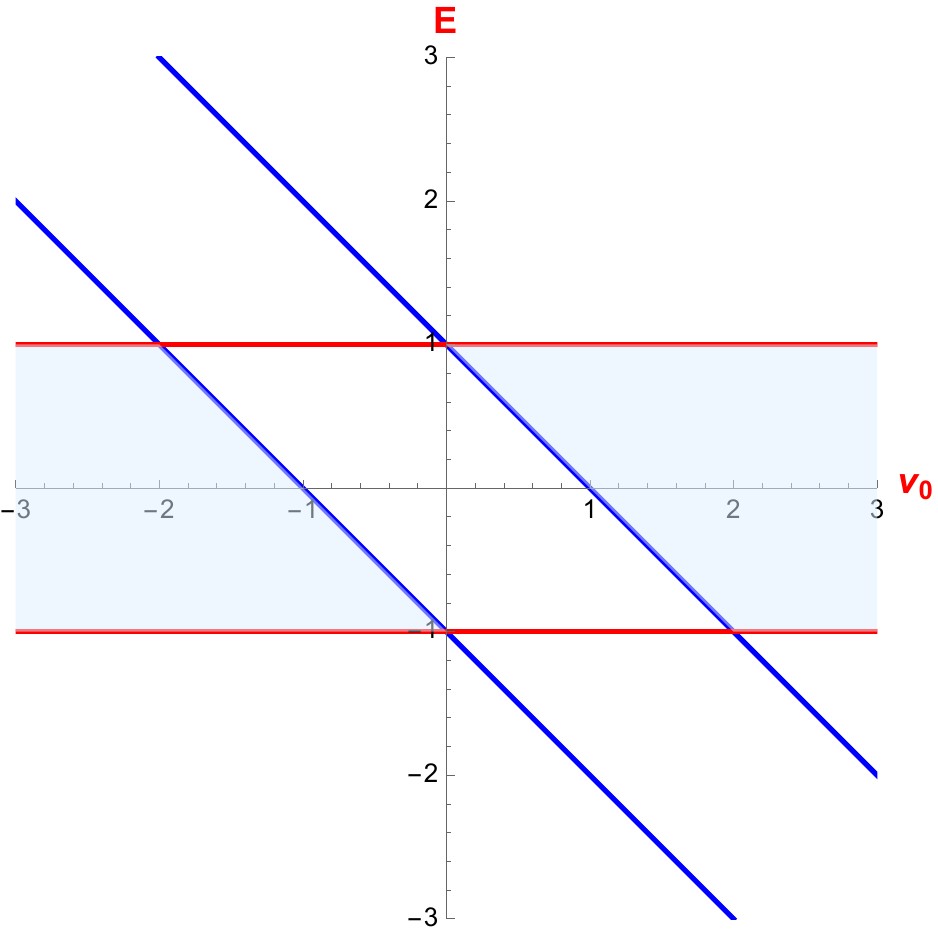}
\caption{\small (left) Regions shaded in blue for bound states in an electric well with $v_0=1$ for 
energy $E$ and $y$-momentum $k$. The red lines corresponds to the first inequality of (\ref{elect1}) and the blue lines to the second inequality.  (right) Regions in the energy $E$-$v_0$ plane for bound states of an electric well with $k=1$, 
according to the same inequalities with the same colour as before. We have also included bound regions for  negative values $k$ and $E$, according to (\ref{elect1}). Notice the $k$-reflection symmetry
(left) and the symmetry $(E,v_0)\to (-E,-v_0)$ (right) of these regions.  \label{fig1}
}
\end{center}
\end{figure}

\begin{itemize}

\item {\it Region II. Bounded motion for $|E|<k$ and $|E+v_0|> |k|$}

In this subsection we describe the classical motion restricted to the waveguide, region II. The particle inside region II may be trapped in such region, if it is not coming from region I as in the above scattering process.
In this case, we have the following values of energy and momenta inside region II: 

i) Since the particle can not go out of region II (it is forbidden in I): $ \  |E|< |k|$. 

ii) If the particle is in region II: $(E+v_0)^2 = (p'_x)^2+ k^2$,
 therefore $|E+v_0|>|k|$. 
 
Thus, $E$ must satisfy the following inequalities (we allow also $E$ and $k$ to be negative) 
\begin{equation}\label{elect1}
\begin{array}{ll}
i)\ \  |E|< |k|,\qquad &ii)\ \  |E+v_0|>|k| 
\end{array}
\end{equation}
Then, the region of parameters satisfying these two inequalities for a particle trapped in region II, is shown in Fig.~\ref{fig1}.
If $k>v_0$, then all the interval $(k-v_0< E< k)$ is positive. If $k<v_0$, a
part of the interval $(k-v_0, k)$ is negative.


\end{itemize}

Thus, as mentioned above, any particle coming from the left is going to pass, or will be scattered, because there is a well and there are no obstacles to continue its trajectory, see Fig.~\ref{figA}, left and center. 

It is also possible that a classical particle be inside the well without enough energy to go out, which means that there can be bound states. This behaviour is represented in Fig.~\ref{figA} (right). The bound states belong to the regions of Fig.~\ref{fig1}, which satisfy  both inequalities (\ref{elect1}). On the left, these regions (in blue) of the $E$-$k$ plane, for a fixed well intensity $v_0$, are shown. The symmetry of these regions in the parameter $k$ (of the  constant momentum) is explicit. On the right, the regions of the plane $E$-$v_0$, for a fixed $k$--value, where the motion is bounded are also displayed in blue color. In this graphic one can appreciate the symmetry under the change $(E,v_0) \to (-E,-v_0)$. 
\bigskip

\noindent
\subsection{The case of a magnetic well}

Next, let us consider the case of a magnetic well under the same point of view  with the following conditions.
The square magnetic well (barrier) is defined by the vector potential where $A(x)$ is given by
\begin{equation}\label{magnetic}
{ A}(x)= 
\left\{\begin{array}{cl}
-a_0,  \ &|x|<1 \,, 
\\[1.5ex]
 0,  \ &|x|>1
 \end{array}\right.
\end{equation}
We have fixed the width again, leading to the same three regions: $I$ for $x<-1$, $II$ for $-1<x<1$ and $III$ for $x>1$.
If $a_0<0$ it will be a barrier, while $a_0>0$ is for a well.

\subsubsection{Magnetic scattering for $k>0$} 

We will consider a particle incoming from the left into the well of region II where $a_0>0$.
Then, we will see the different options for this particle. In this case the sign of the momentum $p_y=k$ is important due to the Lorentz force. If $k>0$ the potential (\ref{magnetic})  acts like a barrier (the Lorentz force push the incident particle out of the well).
\begin{itemize}

\item
{\it Regions I and III.}

The particle initially has energy $E$, and momenta $p_x$ and $p_y$. Since in the $y$-direction the potential  does not change, the momentum $p_y$ will be a constant, $p_y=k$, as in the electric case. Since the potential in the left region is zero $A(x)=0$, we have again the relativistic zero mass energy-momentum relation
(we take $c=1$)
\begin{equation}\label{r1}
E^2= p_x^2 + k^2 \ \implies p_x=\sqrt{E^2-k^2}
\end{equation}

Besides $k>0$ we will also suppose that $p_x>0$ and $E>0$ (we may also consider negative values of $E$), so that once $p_y=k$ is fixed, then from (\ref{r1}) $E \geq k$.
We have the initial ranges of values: 
\[
k\leq E <\infty,\qquad 0 \leq p_x <\infty
\] 

\begin{figure}[h!]\label{energy1}
\begin{center}
\includegraphics[scale=0.35]{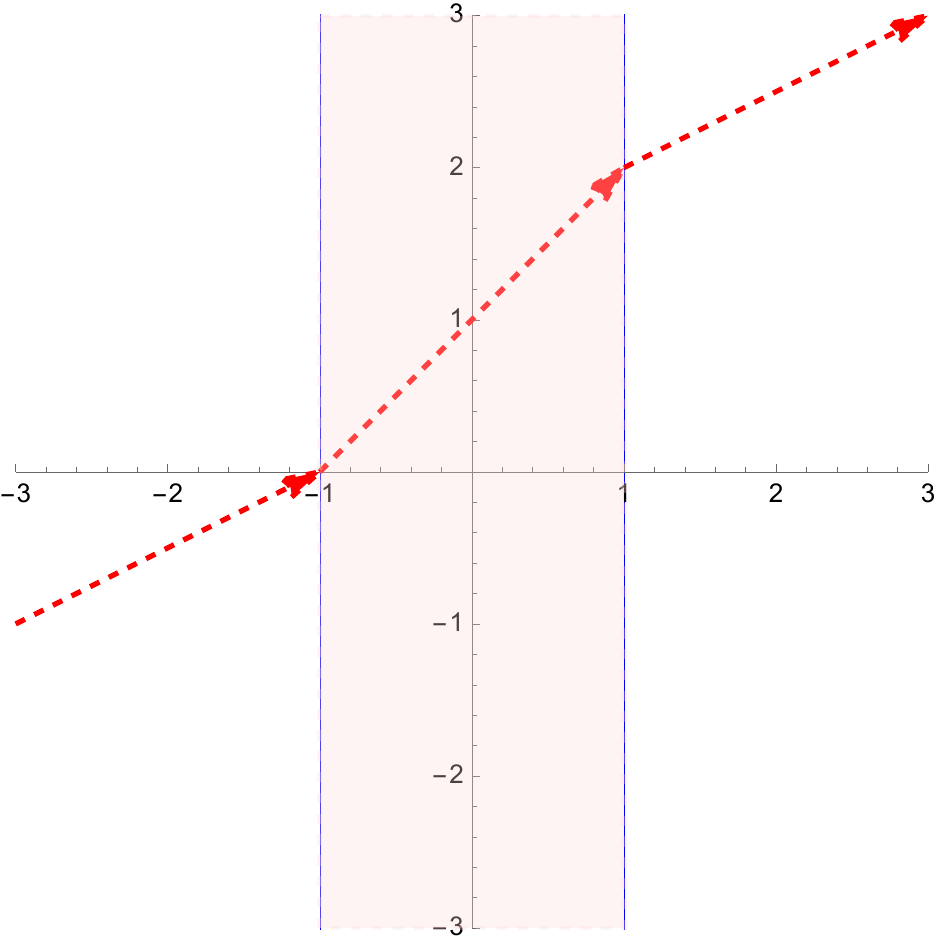}
\includegraphics[scale=0.35]{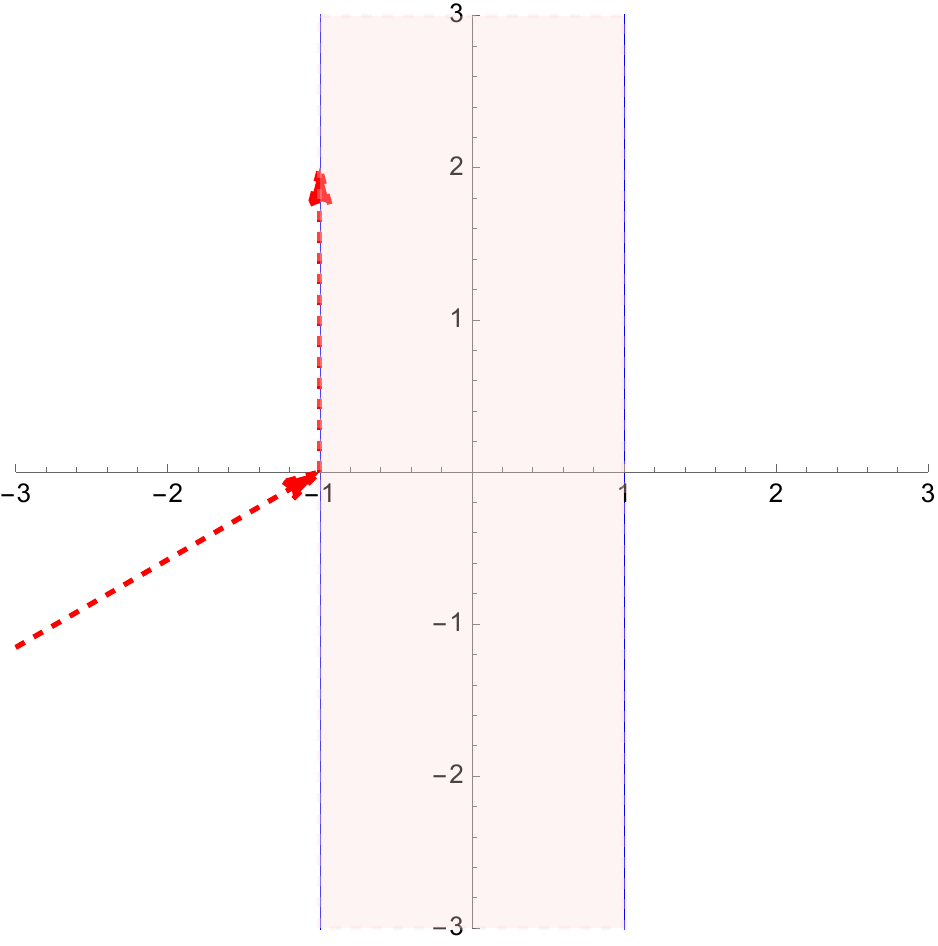}
\includegraphics[scale=0.35]{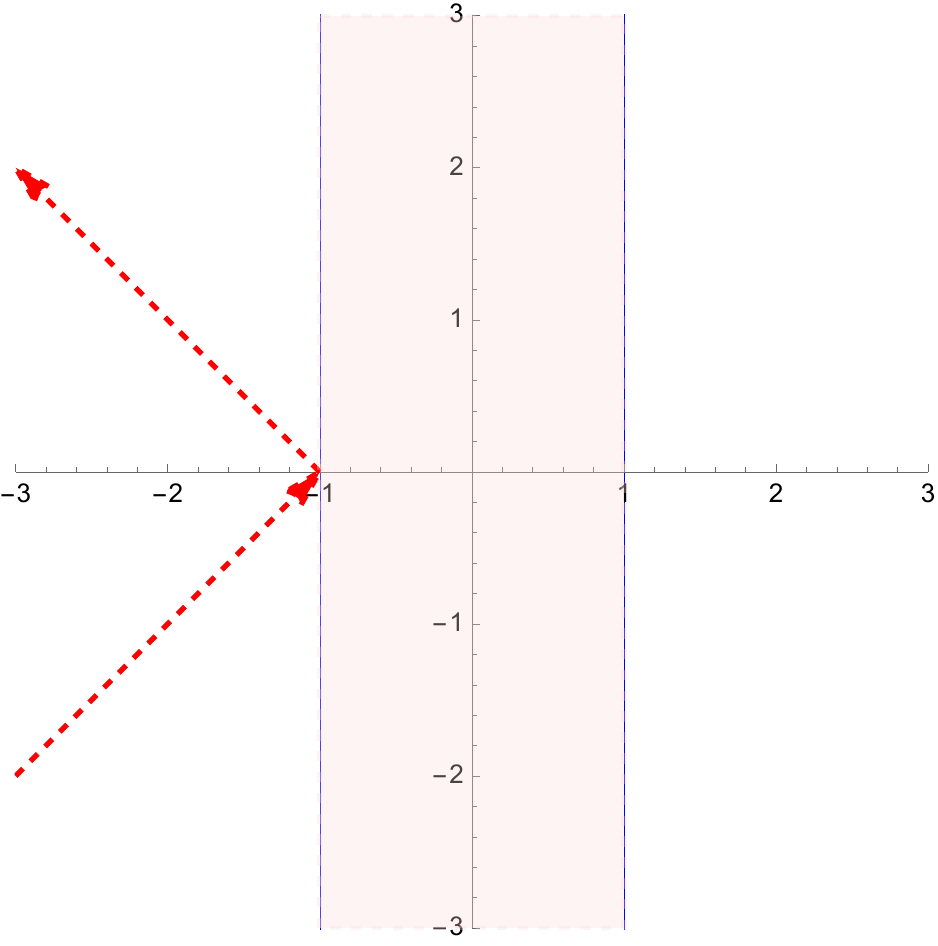}
\caption{\small Trajectories of a particle with $k>0$ in a repulsive magnetic well with configuration $k=1$, $a_0=1$ for different values of the incident momenta: (a) $p_x=2$ (scattering) (b) $p_x=p_{\rm min}=1/\sqrt{3}$ (limit momentum) and (c) $p_x<p_{\rm min}$ with total reflection.  \label{magnetic1}}
\end{center}
\end{figure}

\item
{\it Region II. $k>0$. (Fig.~\ref{magnetic1})
}

Once the particle passes to  region II the energy is the same as in region I, while the canonical momentum  in the $y$ direction ($p_y= k$) has to adapt to the existing magnetic potential.
In this case $p_y$ will change to $p_y-A$, according to the minimal coupling rule for electromagnetic interactions. Therefore, in region II we will have the value  $p'_y=k+a_0$.
Since $a_0>0$, then this value of  $p'_y$ inside the well is also positive and greater than $p_y$:
\[
 p_y=k,\qquad p'_y=k+a_0 
\] 
The other component of the momentum (in region II)  $p'_x$ is obtained from the energy-momenta relation. In this case $p'_x$  is less than $p_x$ (momentum in region I), 
\begin{equation}\label{r3}
E^2= (p'_x)^2 + (k+a_0)^2 \ \implies p'_x=\sqrt{E^2-(k+a_0)^2}<p_x
\end{equation}
\smallskip

%
%

Recall that the initial energy $E$ is greater than $k$: $E\geq k$ (because the particle comes from region I). But due to the last inequality, if $p'_x$ and $p_x$ are positive,
\begin{equation}\label{scatmag}
|E|\geq |k+a_0|
\end{equation}
Then, if the particle pass to region II, the minimum value of the energy is  $E= k+a_0$ and therefore:
\[
p_{\rm min} = \sqrt{(k+a_0)^2-k^2}
\]
Then, in order to have a scattering crossing the magnetic well, the incident angle
$\alpha$ must be less than a maximum angle $\alpha_{\rm max}$, 
\begin{equation}\label{almax}
\sin \alpha = \frac{k}{E} \leq  \frac{k}{k+a_0} = \sin \alpha_{\rm max}
\end{equation}
This is the maximum incident angle for a scattering. Beyond this value of $\alpha_{\rm max}$ (for a fixed $k$), the incident particle will reflect on the boundary of the well.
In summary, there will be scattering (the particle will pass to region II and then to III) only if the incoming energy 
satifies the inequality (\ref{scatmag}).

\end{itemize}
%
%
%

\bigskip

Next, we will consider the case of opposite sign $k= k^-$ (where $k^-$ is negative), then the Lorentz force push the particle into region II, giving rise to three different situations.

\subsubsection{Scattering and bound states for $k^-<0$} 
In this case there will be scattering, bound or reflection depending on the values of $k^- + a_0$.

\begin{figure}[h!]\label{energym2}
\begin{center}
\includegraphics[scale=0.35]{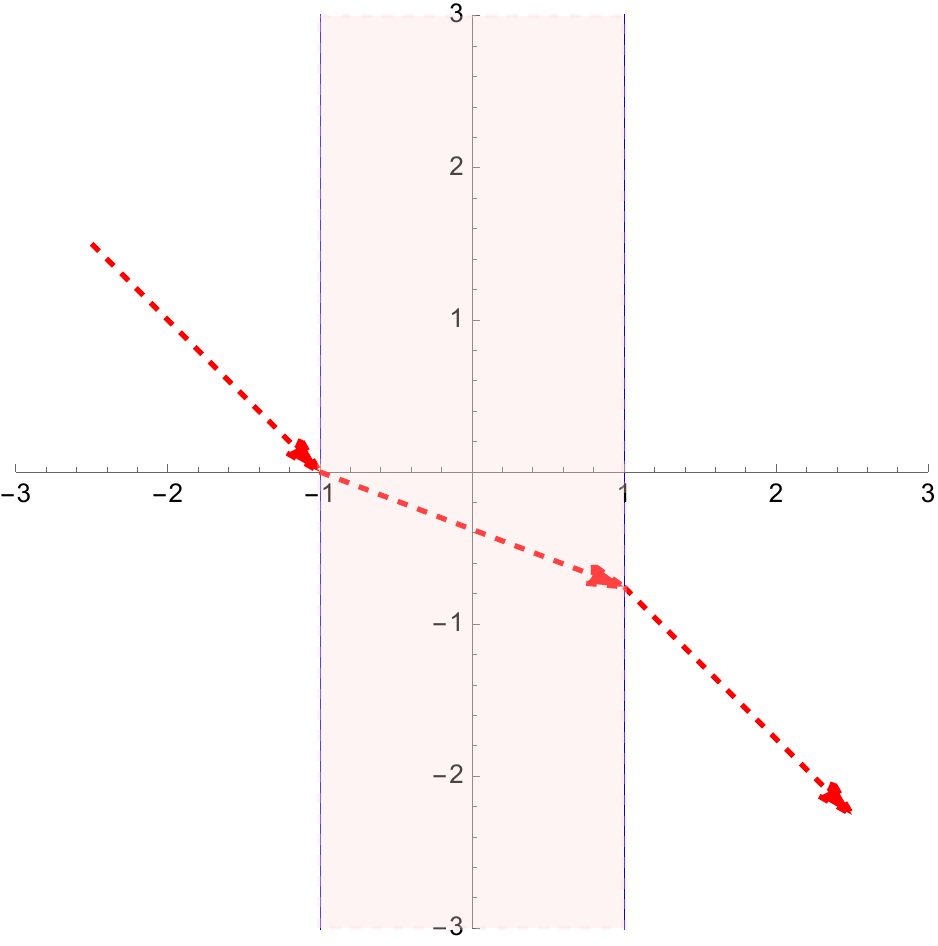}
\includegraphics[scale=0.35]{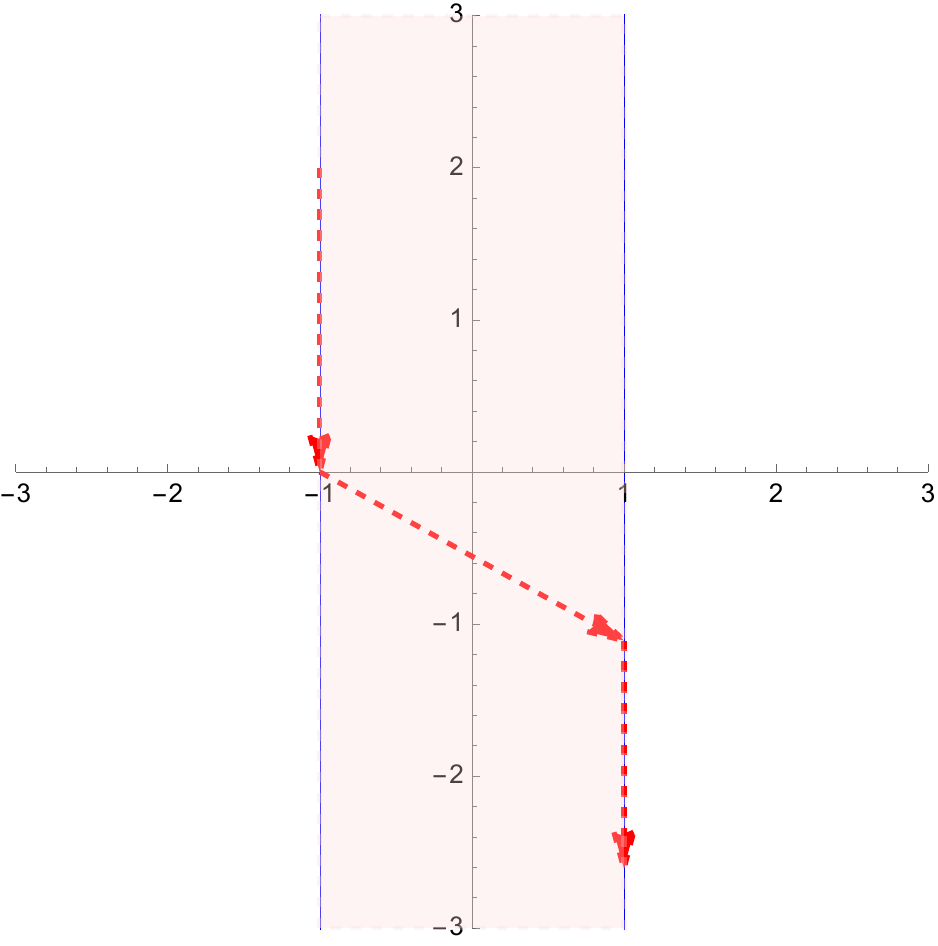}
\includegraphics[scale=0.35]{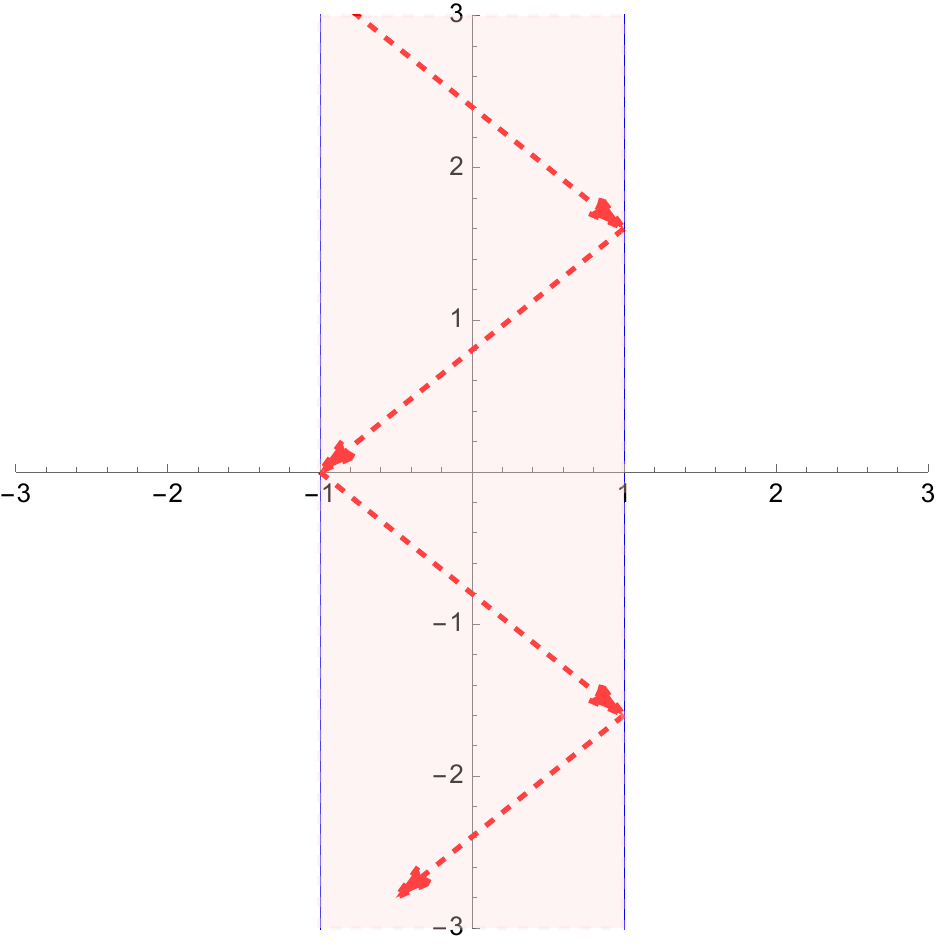}
\caption{\small Trajectories of a particle $k^-<0$ in a magnetic attractive well
($k^- + a_0<0$) with $k^-=-2$ in a magnetic well $a_0=1$ for (a) $p_x=2$, $E=\sqrt{8}$ (left),  (b) $p_x=0$, $E= 2$ (center) and (c) $1<E<2$ for a trapped motion (right).  \label{magnetic2}}
\end{center}
\end{figure}

\begin{figure}[h!]\label{energym3}
\begin{center}
\includegraphics[scale=0.35]{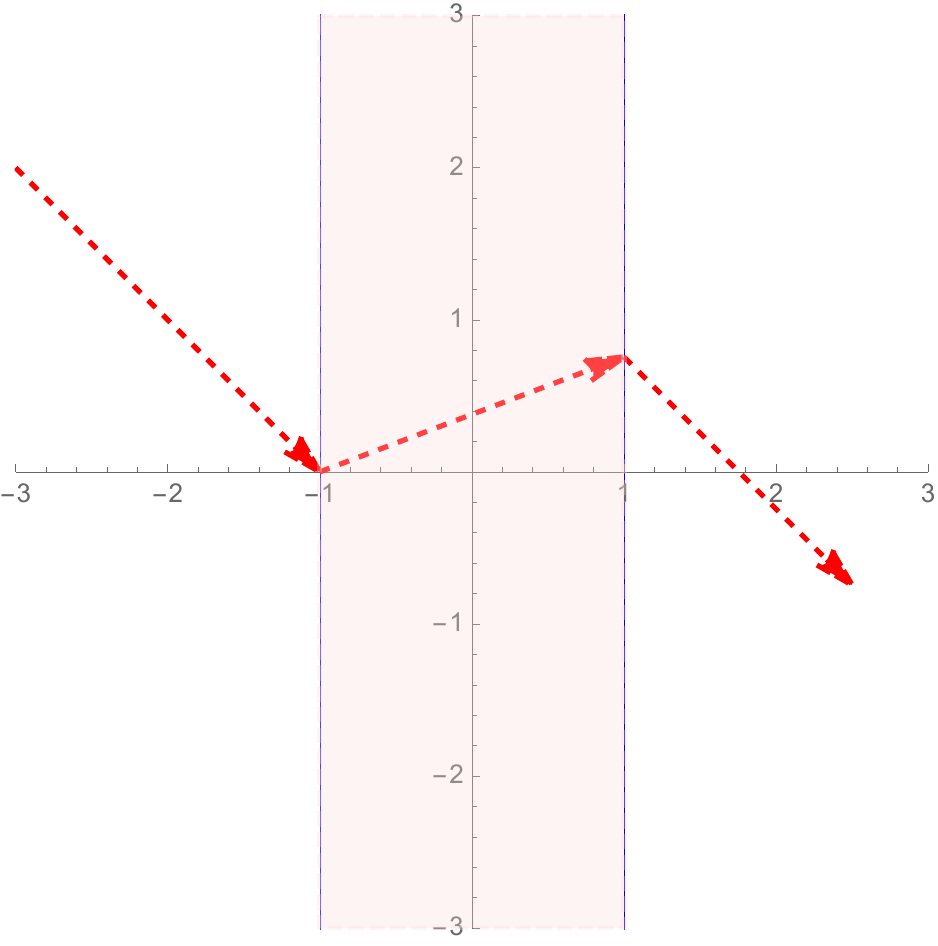}
\includegraphics[scale=0.35]{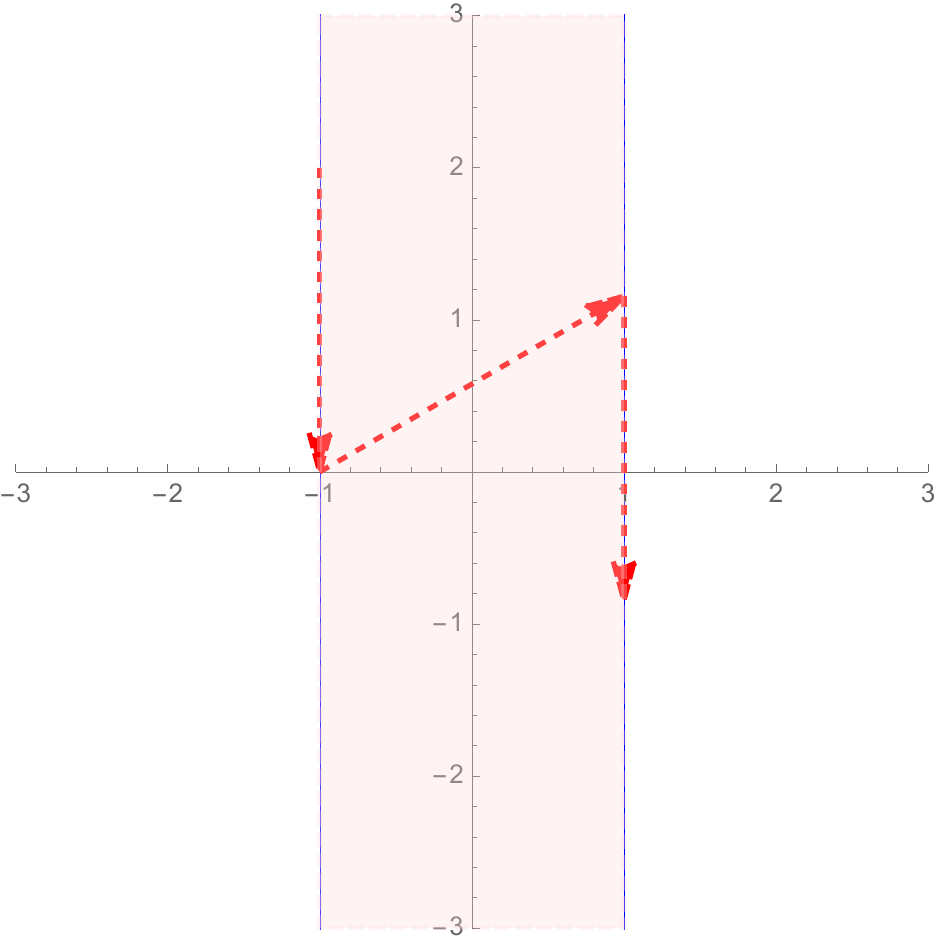}
\includegraphics[scale=0.35]{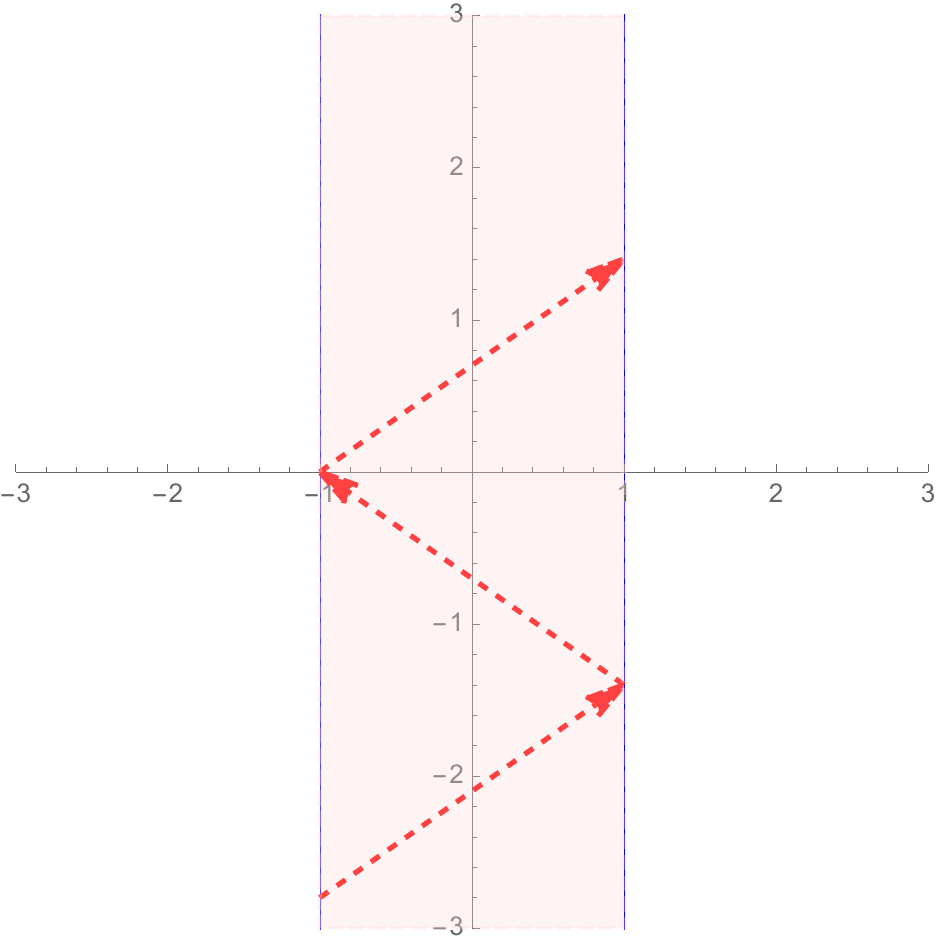}
\caption{\small Trajectories of a particle $k^-<0$ in a magnetic attractive well
($k^- + a_0<0$) with $k^-=-2$ in a magnetic well $a_0=3$ for (a) $p_x=2$, $E=\sqrt{8}$
(left),  (b) $p_x=0$, $E= 2$ (center) and (c) $1<E<2$ for a trapped motion (right).  \label{magnetic3}}
\end{center}
\end{figure}

\begin{figure}[h!]\label{energym4}
\begin{center}
\includegraphics[scale=0.35]{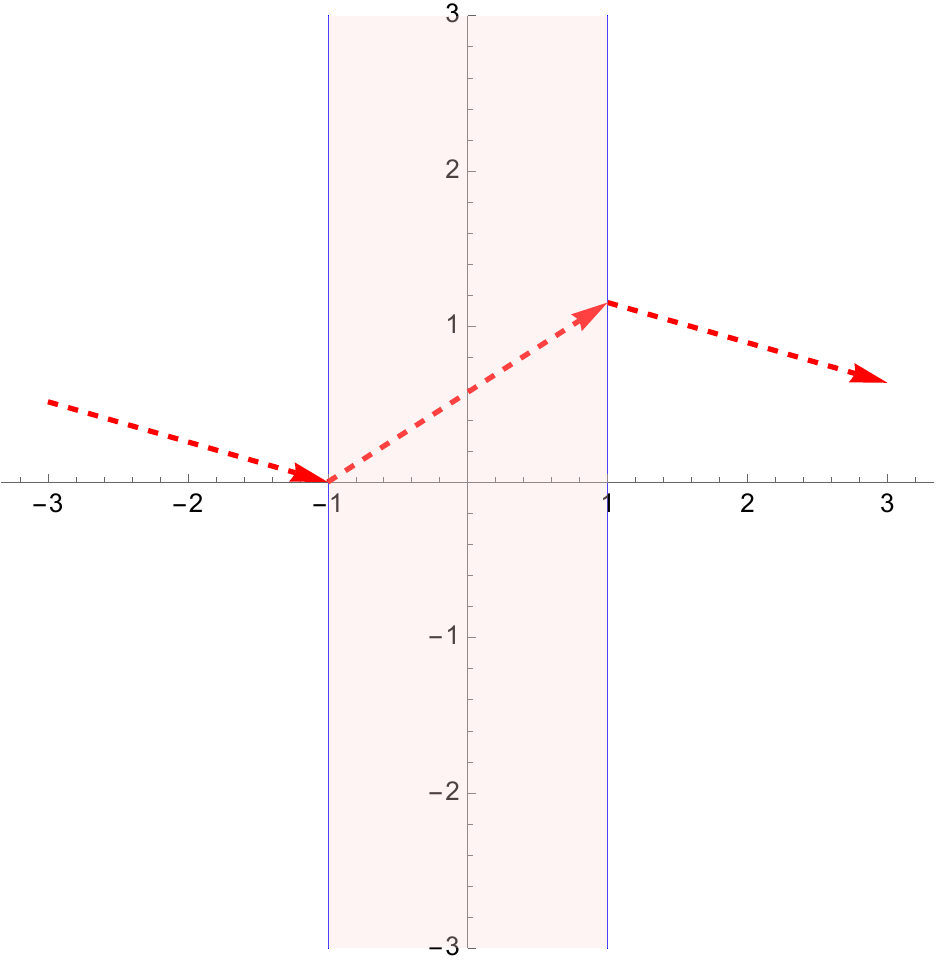}
\includegraphics[scale=0.35]{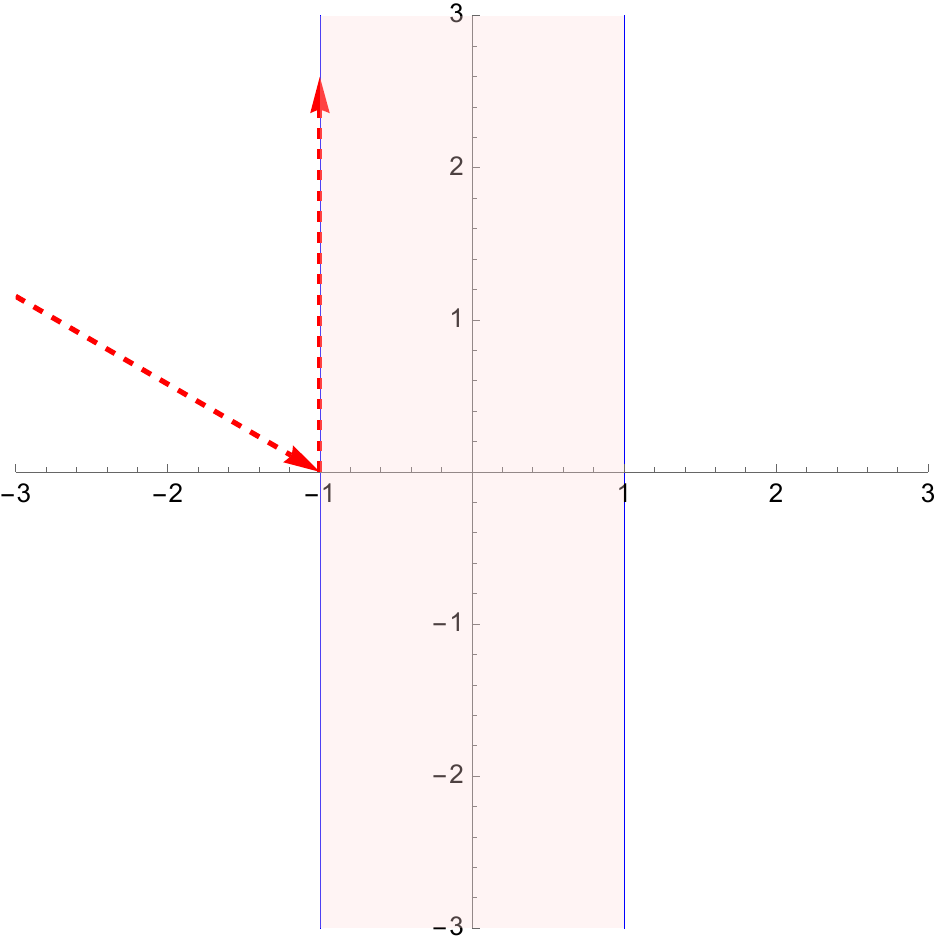}
\includegraphics[scale=0.35]{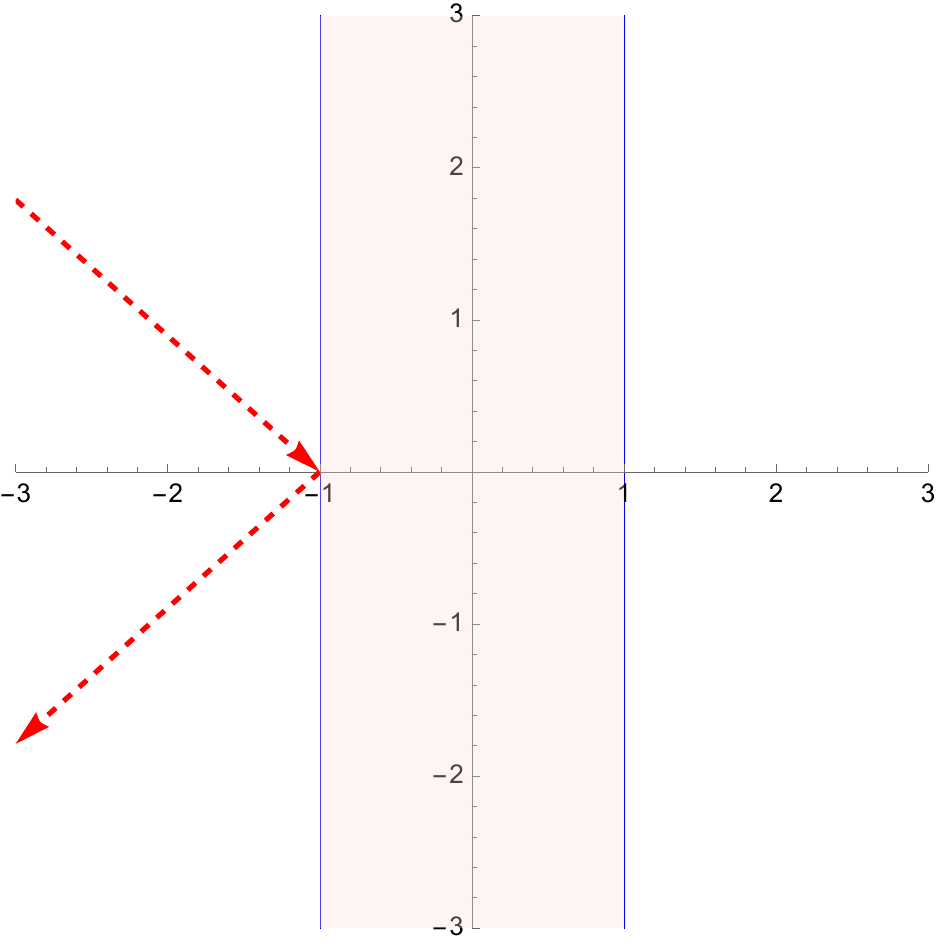}
\caption{\small Trajectories of a particle $k^-<0$ in a magnetic attractive well
($|k^- + a_0|>|k^-|$). We choose the values $k^-=-2$ and  $a_0=6$ for (a) $E=8$ (left),  (b), $E= 4$ (center) and (c) $E=3$ (right).  \label{magnetic4}}
\end{center}
\end{figure}

If for a well $a_0>0$, the $k$-momentum is negative $k^-<0$ then 
\[
p'_y=k^-+a_0= -|k^-|+|a_0|
\] 
The other component of the momentum is $p'_x$. There are three possibilities:

\begin{itemize}
\item[(i)]  $k^-+a_0<0$. {\it(Fig.~\ref{magnetic2})}

This means that $p'_y= k^-+ a_0 < |k^-|$ and therefore $p'_x> p_x$.
If $p_x=0$, $E=|k^-|$ then $p'_x= \sqrt{(k^-)^2- (k^-+a_0)^2}= p'_{\rm min}$ is the minimum value of $p'_x$ in the scattering process. The angle $\sin \alpha'_{\rm max}=\frac{k^-+a_0}{|k^-|}$ is the maximum angle inside the well for scattering states. In conclusion there will always be scattering where the particle will be crossing the barrier ($E^2>k^2$).
The  trajectory is characterized by the negative incident angle $
\sin \alpha=\frac{k^-}{E}
$,  
and the negative refraction angle $\sin \alpha'=\frac{k^-+a_0}{E}$. Here, the negative signs of $\alpha$, $\alpha'$  mean that trajectories in all the regions go in the same negative $y$-sense, in this case with
$|\alpha|>|\alpha'|$ (see Fig.~\ref{magnetic2}).

Bound states exist for 
\begin{equation}\label{b1}
|k^-+a_0|<E<|k^-|,
\end{equation}
and the angle of the bound trajectories are greater than  $\alpha'_{\rm max}$:
\begin{equation}\label{b1b}
|\sin \alpha'| = \frac{|k^-+a_0|}{|E|}>\frac{|k^-+a_0|}{|k^-|}
\end{equation}

\item[(ii)]  $0<k^-+a_0<|k^-|$. {\it (Fig.~\ref{magnetic3})}

In this case $p'_x> p_x$.
If $p_x=0$, $E=|k^-|$, then $p'_x= \sqrt{(k^-)^2- (k^-+a_0)^2}= p'_{\rm min}$ is the minimum value of $p'_x$ in the scattering process. In this case $\alpha$ is negative, while $\alpha'$ positive and $|\alpha|>|\alpha'|$.
This means that in this situation the refractive index is negative.
 As in the previous case the angle $\sin \alpha'_{\rm max}=\frac{k^-+a_0}{|k^-|}$ is the maximum angle inside the well for scattering states, but with positive sign.
Bound states exist for 
\begin{equation}\label{b2}
k^-+a_0<E<|k^-|
\end{equation}
characterized by the angle  $\sin \alpha'=\frac{k^-+a_0}{|E|}$, with opposite sign to the previous case.

\item[(iii)]  $|k^-| <k^-+a_0 $. {\it (Fig.~\ref{magnetic4})}

Then,  we have a very deep well without bound states, $p'_x< p_x$.
Only if $E>k^-+a_0$ there will be scattering, with angles of opposite sign:
\[
\sin \alpha =\frac{k^-}{E}<0,\quad \sin \alpha' =\frac{k^-+a_0}{E}>0, \quad |\alpha'|>|\alpha|
\]
This magnetic waveguide will also correspond to a negative index medium. 
If $|k^-|<E< k^-+a_0$ then  there is a total reflection. The angle $\sin \alpha_{\rm max}= k^-/(k^- + a_0)$ gives the maximum angle for scattering. Beyond that value, there is reflection.
No bound states  exist.

\end{itemize}

\begin{figure}[h!]\label{energy1}
\begin{center}
\includegraphics[scale=0.35]{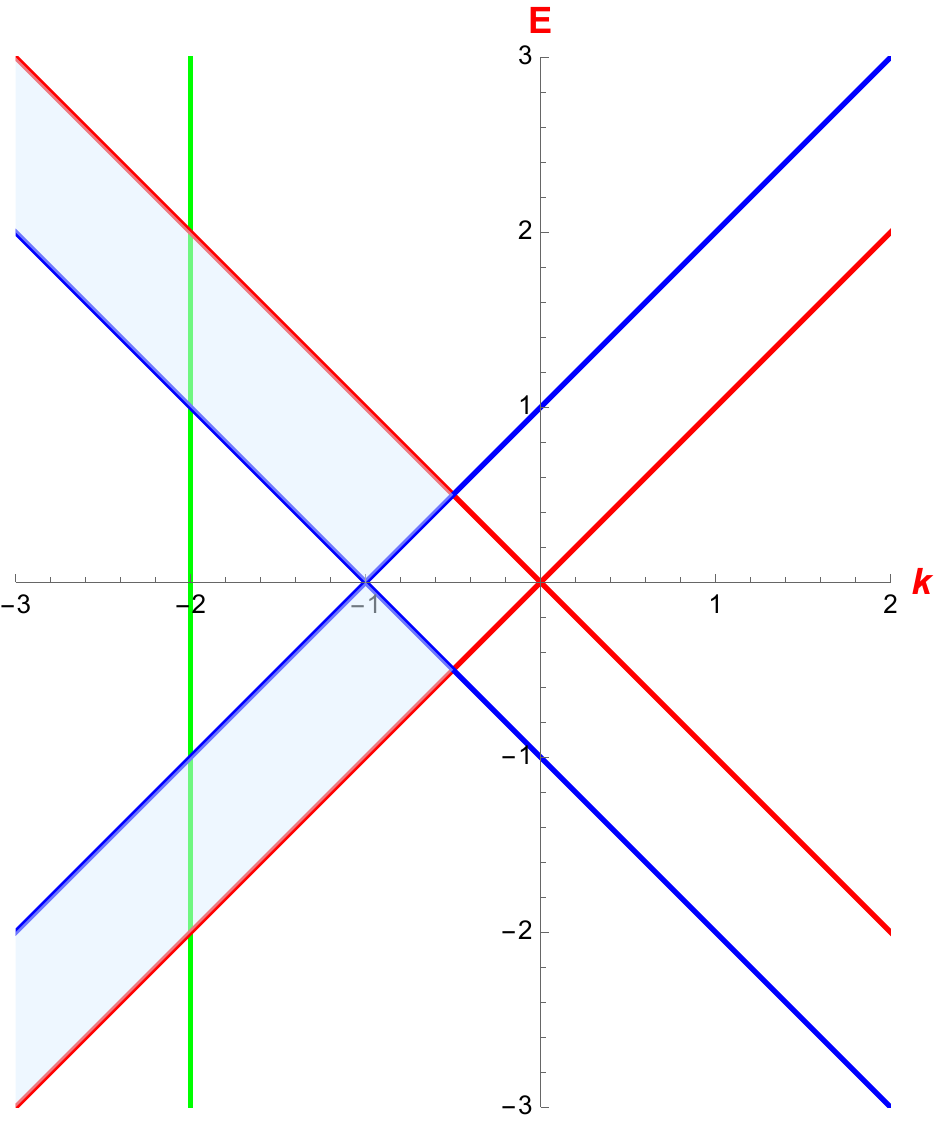}
\qquad
\includegraphics[scale=0.35]{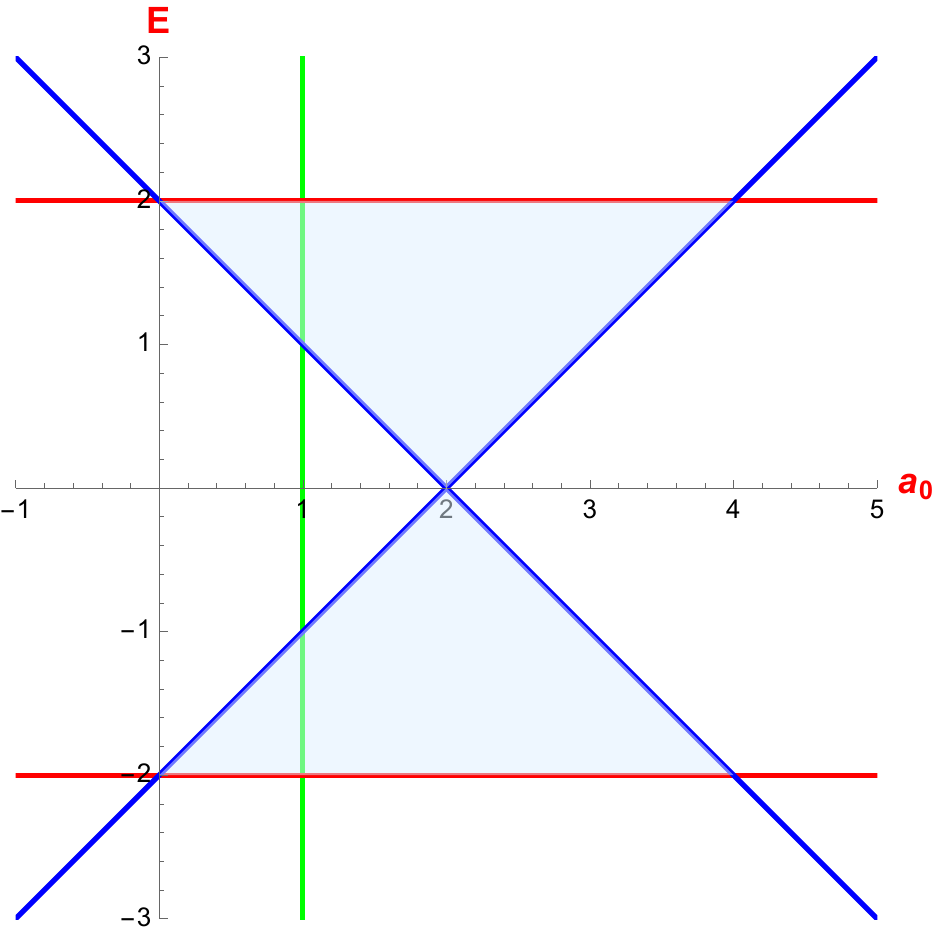}
\caption{\small (left) Regions of the plane energy-momentum $(E,k)$ for the bound states of a magnetic well with depth $a_0=1$. (right) Regions of the plane energy-depth $(E,a_0)$ for bound states for a fixed momentum $k=-2$.   \label{7ab}}
\end{center}
\end{figure}

In summary (see also \cite{peeters10}), there are three cases for negative momentum $k^-$: i) for $k^-<k^-+a_0<0$ where the bound states propagate in the negative $y$-direction; ii) for $|k^-|>k^-+a_0>0$
where the bound (or refracted in region II) states propagate in the positive $y$-direction; and iii) 
for $|k^-|<k^-+a_0$, where the magnetic well is so intense that no bound states exist. This surprising behaviour of bound states as a function of magnetic intensity $a_0$ can be seen in Fig.~\ref{7ab} (right) where the regions of bound states have a (symmetric) triangular shape. These regions correspond to inequalities (\ref{b1}), where $k^-$ is fixed. There is  a high contrast to the electric bound states as a function of the depth of the potential in Fig.~\ref{fig1}, where one can find bound states for any high value of $v_0$. 

On the left graphic of Fig.~\ref{7ab} it is shown the regions of bound states as a function of the vertical momentum $k$. This corresponds to inequalities (\ref{b2}), once $a_0$ is fixed. In this graphic, we see that only for $k<0$ there are bound motions and there is a symmetry of bound states for positive and negative energies $E$. 

%
%

\medskip

\sect{Quantum electric and magnetic wells}

In the following we will examine the solutions of the quantum problem of electric and magnetic waveguides, in the context of graphene quasiparticles. We expect the following situations.

i) In the range of energies of total classical scattering we may find partial scattering with a non null reflection amplitude. Total scattering (total transmision) will take place for discrete values of the relevant parameter (here we will take the angle $\alpha$ or the energy $E$).

ii) In the range of energies  of bound or trapped classical states there will be a set of discrete values of energy for true bound quantum states.

iii) In the case of energies with total reflection we may find some quantum states of partial tunnelling, allowing the transmission to the outgoing region.

iv) There could even exist bound quantum states in forbidden classical regions: scattering, trapped or reflection states. This is the case of edge states \cite{lado15}. We will see that the  magnetic bound states will allow two of them to have an edge  character. This fact is quite important because edge states have special properties and  there is only some special configurations allowing for such states.

\subsection{Quantum magnetic well}

Using the vector  potential (\ref{magnetic}) and the spinor form
\begin{equation}\label{psi}
\Psi({\bf x})= e^{i k\, y} e^{i k_x x}
\left(\begin{array}{l}
\phi_1 
\\[1.5ex]
\phi_2  
 \end{array}\right)
\end{equation}
where $\phi_1, \phi_2$ are constants, in (\ref{dirac2}), we can obtain the solutions of (\ref{dirac2}) in the three regions: region I ($x<-1$), region II ($-1<x<1$), and region III ($x>1$).

\subsubsection{Scattering}
We will start by the scattering states of a magnetic waveguide.

\begin{itemize}
\item {\it Region I.} ($x<-1$)

Consider the eigenvalue problem
\begin{equation}\label{psi1}
\left(\begin{array}{cc}
0   & (-i+k_x)  
\\[1.5ex]
(ik+k_x)  &   0
 \end{array}\right)\left(\begin{array}{l}
\phi_1 
\\[1.5ex]
\phi_2 
 \end{array}\right)=E
\left(\begin{array}{l}
\phi_1 
\\[1.5ex]
\phi_2 
 \end{array}\right)
\end{equation}

From this matrix equation, we have two equations for $\psi_1$ and $\psi_2$:
\begin{equation}\label{psi1}
(-i+k_x)\phi_2=E \phi_1, \qquad (ik+k_x)\phi_1=E \phi_2  
\end{equation}
where $k_x$ and $k$ are constants.  The consistency of both equations give
\begin{equation}\label{ee}
E^2 = k_x^2 + k^2\,,\qquad E_\pm = \pm \sqrt{k_x^2 + k^2}
\end{equation}
The two signs of $E_\pm$ correspond to positive or negative real energies.
If we take $\phi_1=1$, 
\begin{equation}\label{phi2}
\phi_2=\frac{(k_x+ik)}{E_{\pm}}
\end{equation}
Since for each energy $E_\pm$ we have $k_x=\pm\sqrt{E^2-k^2}$, then we get the solutions
\begin{equation}\label{psi1phiX}
\Psi_I^{\pm}(E_\pm,{\bf x})= e^{i k y} e^{\pm i k_x x}
\left(\begin{array}{c}
1
\\[1.5ex]
\frac{(\pm k_x+ik)}{E_{\pm}}
 \end{array}\right)=
 e^{i k y} e^{\pm i k_x x}
\left(\begin{array}{c}
1
\\[1.5ex]
e^{i\alpha_{\rm i/o}}
 \end{array}\right),
 \qquad  
 k_x=\sqrt{E^2-k^2}
\end{equation}
where ${\bf x} = (x,y)$ and the incident angle is $\alpha_{\rm i}=\arctan \frac{ k}{k_x}$ and  $\alpha_{\rm o}=\arctan \frac{- k}{k_x}$
the reflected angle. Thus, the general solution is
\begin{equation}\label{psi1phiI}
\Psi_I(E_\pm,{\bf x})= \alpha_1\,\Psi_I^+(E_\pm,{\bf x})
+ \alpha_2\,\Psi_I^-(E_\pm,{\bf x})
\end{equation}
where $\alpha_1, \alpha_2$ are arbitrary constants.

\item {\it Region II.} ($-1<x<1$)

For this region from the matrix equation taking into account that, due to the magnetic potential, $-i\partial_y$ is replaced by $-i\partial_y+a_0$ (see eq.~(\ref{dirac2})), we get the equations for $\phi_1$ and $\phi_2$:
\begin{equation}\label{psi1}
(-i(k+a_0)+k'_x)\phi_2=E \phi_1, \qquad (i(k+a_0)+k'_x)\phi_1=E \phi_2  
\end{equation}
Solving them together, we obtain $\phi_1$ and $\phi_2$ and we can write the independent solutions 
$\Psi_{II}^\pm(E_\pm,\bf x)$
\begin{equation}\label{psi2phipmX}
\Psi_{II}^{\pm}(E_\pm,{\bf x})= e^{i k y} e^{\pm i k'_x x}
\left(\begin{array}{c}
1 
\\[1.5ex]
\frac{(\pm k'_x+i(k+a_0))}{E_{\pm}}
 \end{array}\right)\,,\qquad
 k'_x=\sqrt{E^2-(k+a_0)^2}
\end{equation}
The general solution is
\begin{equation}\label{psi2phigen}
\Psi_{II}(E_\pm,{\bf x})= \beta_1 \Psi_{II}^{+}(E_\pm,{\bf x}) +
\beta_2 \Psi_{II}^{-}(E_\pm,{\bf x})
\end{equation}
where $\beta_1,\beta_2$ are arbitrary constants.

\item {\it Region III.}  ($x>1$)

Following the same procedure as above, for this region we have the same solutions of region I:
\begin{equation}\label{psi3phi}
\Psi_{III}^{\pm}({\bf x})= e^{i k y} e^{\pm i k_x x}
\left(\begin{array}{c}
1 
\\[1.5ex]
\frac{(\pm k_x+ik)}{E_{\pm}}
 \end{array}\right)\,,
 \qquad  
 k_x=\pm\sqrt{E^2-k^2}
\end{equation}
\begin{equation}\label{psi1phiIII}
\Psi_{III}(E_\pm,{\bf x})= \gamma_1\,\Psi_{III}^+(E_\pm,{\bf x})
+ \gamma_2\,\Psi_{III}^-(E_\pm,{\bf x})
\end{equation}

where $\gamma_1,  \gamma_2$ are  the constants.
\end{itemize}

{\bf Remark 1.} In order to solve the scattering  problem, we must assume that $k_x$ is real  \begin{equation}\label{macht1}
{k_x}^2=(E_\pm)^2 - (k)^2>0,\qquad |E_\pm| > |k|
\end{equation}
So, there will be  incoming  and outgoing plane waves, in the first and third region, respectively.

{\bf Remark 2.} For the second region we must take into account that
 \begin{equation}\label{macht11}
{k'_x}^2=(E_\pm)^2 - (k+a_0)^2
\end{equation}

$\bullet$ If ${k'}_x^2=(E_\pm)^2 - (k+a_0)^2>0$ or
 \begin{equation}\label{macht1}
|E_\pm|> |k+a_0|
\end{equation}
then the momentum $k'_x$ in region II will be real. Classically this means that the electron will pass the magnetic strip. 

$\bullet$ If ${k'}_x^2=(E_\pm)^2 - (k+a_0)^2<0$ or 
 \begin{equation}\label{macht1}
|E_\pm|< |k+a_0|
\end{equation}
then $k'_x$ will be imaginary. This is going to happen if 
 \begin{equation}\label{macht12a}
|k|<|E_\pm|< |k+a_0|
\end{equation}
This is a case where classically there will be total reflection. 
However, this will not happen in the quantum frame due to tunnelling and we could also expect quantum non classical scattering states for the region determined by the inequalities (\ref{macht12a}).

In this situation, we must take the solutions:
\begin{equation}\label{psi2phipmbXX}
\Psi_{II}^{\pm}(E_\pm,{\bf x})= e^{i k y} e^{\pm k'_x x}
\left(\begin{array}{c}
1 
\\[1.5ex]
\frac{(\mp i k'_x+i(k+a_0))}{E_{\pm}}
 \end{array}\right)\,,\qquad
 k'_x=\sqrt{(k+a_0)^2- E^2}
\end{equation}
Thus, the general solution in region II is
\begin{equation}\label{psi2phigen}
\Psi_{II}(E_\pm,{\bf x})= \beta_1 \Psi_{II}^{+}(E_\pm,{\bf x}) +
\beta_2 \Psi_{II}^{-}(E_\pm,{\bf x})
\end{equation}
where $\beta_1,\beta_2$ are arbitrary constants.

Finally, we consider the following type of scattering solutions:
\begin{equation}\label{psi1g22}
\Psi_{I}({\bf x})= \Psi_{I}^+({\bf x}) + r\, \Psi_{I}^-({\bf x}),\qquad \Psi_{II}({\bf x})= A \Psi_{II}^+({\bf x}) + B \Psi_{II}^-({\bf x}),\qquad \Psi_{III}({\bf x})= t\, \Psi_{III}^+({\bf x})
\end{equation}
where $r$  and $t$ are the reflection and transmission amplitudes, respectively. Depending on the values of the $E, k$ and $a_0$, we may use for the solutions $\Psi_{II}$, the formulas  (\ref{psi2phipmX}) or (\ref{psi2phipmbXX}). These solutions must satisfy the matching conditions at $x=-1$ and $x=1$:
\begin{equation}\label{macht12}
\Psi_{I}(-1)=\Psi_{II}(-1),\qquad \Psi_{II}(1)=\Psi_{III}(1)
\end{equation}

If we have a particle with momenta ${\bf k}=(k_x, k)$ with 
$k_x= \sqrt{E^2-k^2}$ corresponding to energy $E$, then the incident angle (the angle of the incident direction with the normal to the magnetic barrier/well) will be
\begin{equation}\label{incident}
\sin \alpha = \frac{k}{|E|}\ \implies \ |E|= k\, \csc \alpha 
\end{equation}
In this way, we can compute the reflection and transmission coefficients depending on the incident angle (for fixed $k$ and $a_0$)
\begin{equation}\label{macht22}
R(\alpha,k,a_0)= |r(\alpha,k,a_0)|\,, \qquad T(\alpha,k,a_0)= |t(\alpha,k,a_0)|
\end{equation}
They depend on the values $k$ of the $y$-momentum,  the depth $a_0$ of the well and  the incident angle $\alpha$ (or equivalently the energy $E$). For example the transmission coefficient, obtained by finding the solution (\ref{psi1g22}) by solving the matching conditions (\ref{macht12}) in  terms of energy (for the allowed classical scattering) is given by
\begin{equation}
 t(E, k, a_0) = 
\frac{ \exp[-2 i k_x]\, k_x k'_x}{k_x k'_x
       \cos(2 k'_x) - 
     i (E^2 - k (a_0 + k)) \sin(2 k'_x)}
\end{equation}
and for the ``forbiden'' classical scattering is given by
\begin{equation}
 t(E, k, a_0) = 
\frac{ \exp[-2 i k_x] |k'_x|^2k_x }{ |k'_x|^2 k_x
       \cosh(2 |k'_x|) - 
     i (E^2 - k (a_0 + k)) \sinh (2 k'_x)}
\end{equation}

\begin{figure}[h!]
  \centering 
\includegraphics[width=0.45\textwidth]{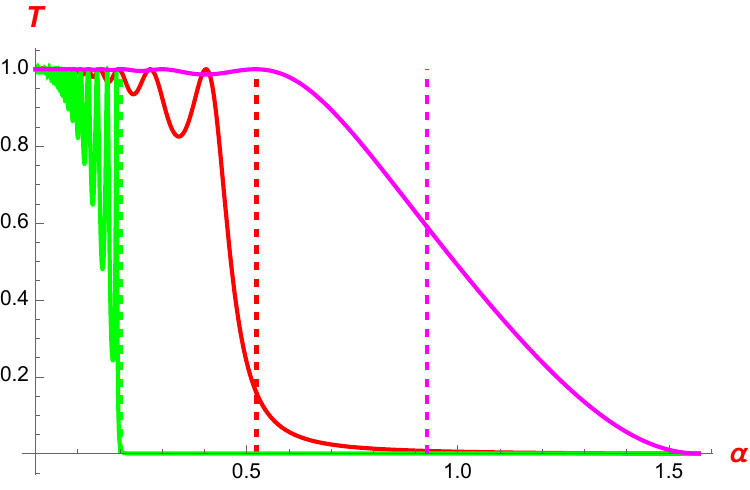}
\quad
\includegraphics[width=0.35\textwidth]{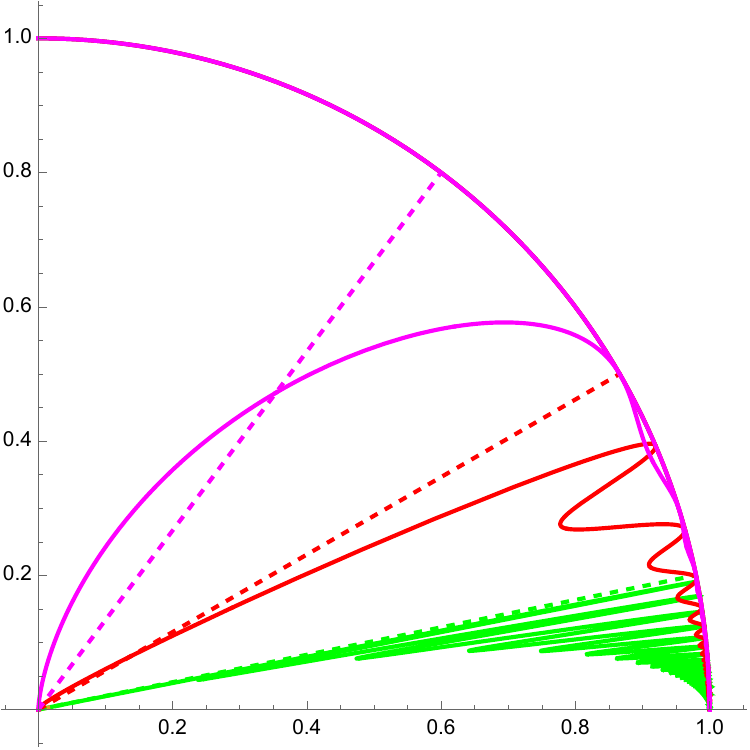}
\caption{(left) Transmission coefficients for $k=1$ and different values of the well depth:  $a_0=4$ (green), $a_0=1$ (in red), $a_0=1/4$ (magenta). (right) The same results in a circular plot of the transmission coefficient $T$ versus the angle of incidence $\alpha$.
The dashed radius (or vertical lines in the left graphic) correspond to the  limit of allowed/non allowed classical scattering given by formula (\ref{almax}). 
}
  \label{figuras2}
\end{figure}

\begin{figure}[h!]
  \centering
\includegraphics[width=0.45\textwidth]{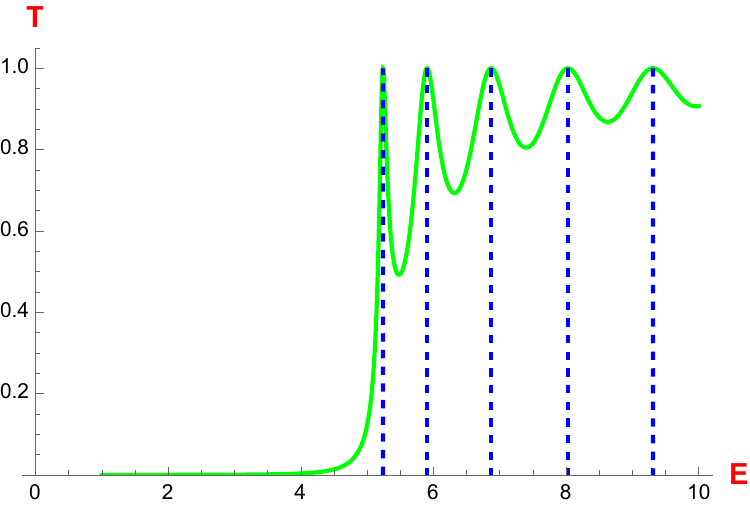}
\caption{Resonances of $T$ (as a function of $E$) for $k=1$, $a_0=4$, appear for the values
$\sin(2 k'_x) = 0$.
}
  \label{figuras3}
\end{figure}

Figs.~\ref{figuras2}-\ref{figuras3} show the transmission coefficient $T= |t|$ of the scattering states for the magnetic well. The resonances $T= |t|=1$ will take place for
the values $\sin(2 k'_x)=0$, i.e., when the phase acquired by the wave in crossing  the well be equal to $n\pi$, $n\in {\mathbb Z}$. This is shown in Fig.~\ref{figuras3}.

\subsubsection{Bound states}

In a similar way, in this subsection we discuss the bound states of a magnetic waveguide.

\begin{itemize}
\item {\it Region I.} ($x<-1$) 

In this region, in order to have bound states, we have the condition: $k^2>E$ and $q_x=\pm\sqrt{k^2- E^2}$. Then the solution (\ref{psi}) has the form:
\begin{equation}\label{psib}
\Psi_I({\bf x})= e^{i k y} e^{q_x x}
\left(\begin{array}{l}
\phi_1 
\\[1.5ex]
\phi_2  
 \end{array}\right)
\end{equation}

\begin{equation}\label{psi1b}
\left(\begin{array}{l}
(-ik-iq_x)\phi_2 
\\[1.5ex]
(ik-i q_x)\phi_1  
 \end{array}\right)=E
\left(\begin{array}{l}
\phi_1 
\\[1.5ex]
\phi_2 
 \end{array}\right)
\end{equation}
From these equations, $E^2=k^2- q_x^2$

\begin{equation}\label{psi1phib}
\Psi_I^{\pm}(E_\pm,{\bf x})= e^{i k y} e^{\pm q_x x}
\left(\begin{array}{c}
1
\\[1.5ex]
\frac{(\mp i q_x+ik)}{E_{\pm}}
 \end{array}\right),
 \qquad  
 q_x= \sqrt{k^2-E^2}
\end{equation}
 The general solution in region I is
\begin{equation}\label{psi1phiIb}
\Psi_I(E_\pm,{\bf x})= \alpha_1\,\Psi_I^+(E_\pm,{\bf x})
+ \alpha_2\,\Psi_I^-(E_\pm,{\bf x})
\end{equation}
We also have the inequality
\begin{equation}\label{l1}
|E|<|k|
\end{equation}

\item {\it Region II.} ($-1<x<1$)

For this region, we get two equations for $\phi_1$ and $\phi_2$:
\begin{equation}\label{psi1b}
(-i(k+a_0)+k'_x)\phi_2=E \phi_1, \qquad (i(k+a_0)+k'_x)\phi_1=E \phi_2  
\end{equation}
Solving them together, we obtain $\phi_1$ and $\phi_2$ and we can write the independent solutions 
$\Psi_{II}^\pm(E_\pm,\bf x)$
\begin{equation}\label{psi2phipmb}
\Psi_{II}^{\pm}(E_\pm,{\bf x})= e^{i k y} e^{\pm i k'_x x}
\left(\begin{array}{c}
1 
\\[1.5ex]
\frac{\pm k'_x+i(k+a_0)}{E_{\pm}}
 \end{array}\right)\,,\qquad
 k'_x=\pm\sqrt{E^2-(k+a_0)^2}
\end{equation}
Then, we have the inequality
\begin{equation}\label{l2}
|E|>|k+a_0|
\end{equation}

Thus, the general solution is
\begin{equation}\label{psi2phigenb}
\Psi_{II}(E_\pm,{\bf x})= \beta_1 \Psi_{II}^{+}(E_\pm,{\bf x}) +
\beta_2 \Psi_{II}^{-}(E_\pm,{\bf x})
\end{equation}
where $\beta_1,\beta_2$ are arbitrary constants.

\item {\it Region III.}  ($x>1$)

Following the same procedure as above, for this region we have the same solutions of region I:
\begin{equation}\label{psi3phib}
\Psi_{III}^{\pm}({\bf x})= e^{i k y} e^{\pm q_x x}
\left(\begin{array}{c}
1 
\\[1.5ex]
\frac{(\mp i q_x+ik)}{E_{\pm}}
 \end{array}\right)\,,
 \qquad  
 q_x=\pm\sqrt{k^2-e^2}
\end{equation}
\begin{equation}\label{psi1phiIIIb}
\Psi_{III}(E_\pm,{\bf x})= \gamma_1\,\Psi_{III}^+(E_\pm,{\bf x})
+ \gamma_2\,\Psi_{III}^-(E_\pm,{\bf x})
\end{equation}

where $\gamma_1,  \gamma_2$ are  the constants.

\end{itemize}

In order to get the bound state  solution, we must assume that $q_x$ for the region I is real  \begin{equation}\label{macht1b}
{q_x}^2=k^2 - (E_\pm)^2>0,\qquad |E_\pm| < |k|
\end{equation}

Next, in  region II we have also two possibilities:
 \begin{equation}\label{macht11b}
{k'}_x^2=(E_\pm)^2 - (k+a_0)^2>0,\quad {\rm or}  \quad {q'}_x^2=(E_\pm)^2 - (k+a_0)^2<0
\end{equation} 
The first possibility gives us the standard bound states and the second one leads to edge states. 
Then, we consider the following solutions for the bound problem:
\begin{equation}\label{psi1gb}
\Psi_{I}({\bf x})= A \Psi_{I}^+({\bf x}),\qquad \Psi_{II}({\bf x})= b \Psi_{II}^+({\bf x}) + C \Psi_{II}^-({\bf x}),\qquad \Psi_{III}({\bf x})= D\, \Psi_{III}^-({\bf x})
\end{equation}
where $A, B, C, D$  are constants. Depending on the values of the $E, k$ and $a_0$, we can use for the solutions $\Psi_{II}$ either standard bound   solutions (\ref{psi2phipmb}) or edge states:
\begin{equation}\label{psi2phipmbe}
\Psi_{II}^{\pm}(E_\pm,{\bf x})= e^{i k y} e^{\pm |q'_x| x}
\left(\begin{array}{c}
1 
\\[1.5ex]
\frac{\mp i |q'_x|+i(k+a_0)}{E_{\pm}}
 \end{array}\right)\,,\qquad
| q'_x|=\pm\sqrt{(k+a_0)^2-E^2}
\end{equation}

 These solutions must satisfy the matching conditions at $x=-1$ and $x=1$:
\begin{equation}\label{macht2}
\Psi_{I}(-1)=\Psi_{II}(-1),\qquad \Psi_{II}(1)=\Psi_{III}(1)
\end{equation}

\begin{figure}[h!]
  \centering
\includegraphics[width=0.4\textwidth]{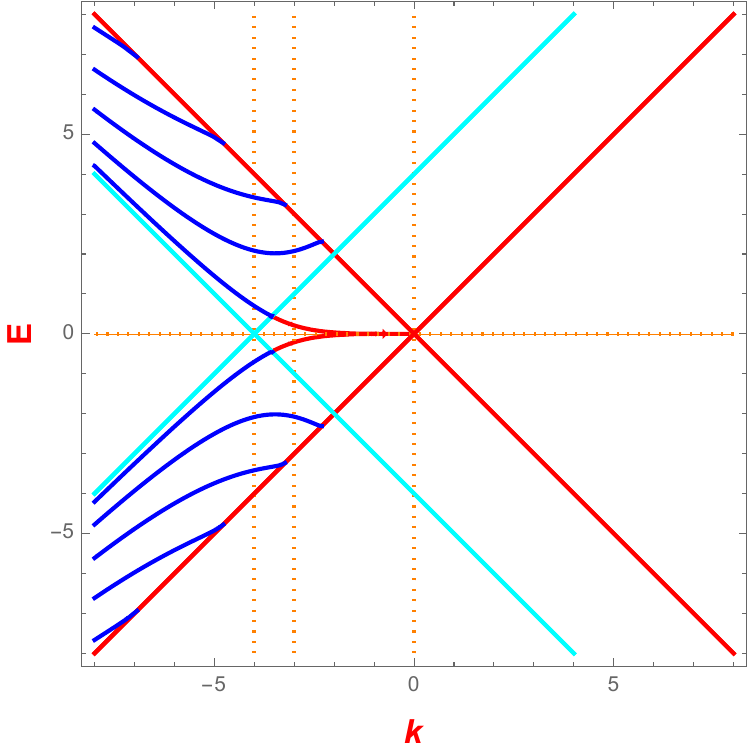}
\quad 
\includegraphics[width=0.4\textwidth]{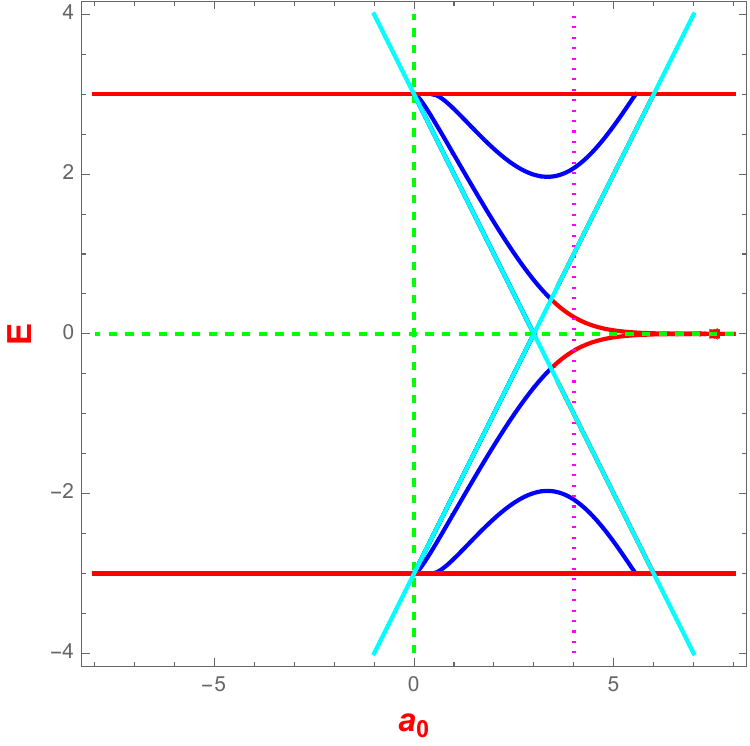}
\caption{(left) Spectral magnetic curves in the energy-$k$ plane $(E,k)$, for $a_0=4$. 
(right) Spectral curves in the  plane momentum-depth $(E,a_0)$, for $k=-3$ of the magnetic well. Red straight lines correspond to inequality (\ref{l1}) and cyan ones to (\ref{l2}). Spectral curves in red out of the classical region correspond to edge states.
}
  \label{10ab}
\end{figure}

The resulting spectrum of bound states is shown in Fig.~\ref{10ab} (see also the discussion in \cite{peeters09}). In the left graphic the parameter of the magnetic potential $a_0=4$ has been fixed. This graphic represent the energy eigenvalues as a function of the vertical momentum $k$. The spectrum are in blue curves inside the classical sector, and outside the classical region bound spectral curves in red correspond to ground edge states. 

If $k$ is fixed, so that the energy depends on the magnetic intensity $a_0$ the corresponding spectral curves are shown in the right of Fig.~\ref{10ab}. This type of magnetic spectrum is completely different to the corresponding electric case, as we will see later. The spectral curves belong to a triangular region, except for the edge states (in red) that are in the exterior of the classical region of  Fig.~\ref{10ab} (right).

\begin{figure}[h!]
  \centering
\includegraphics[width=0.35\textwidth]{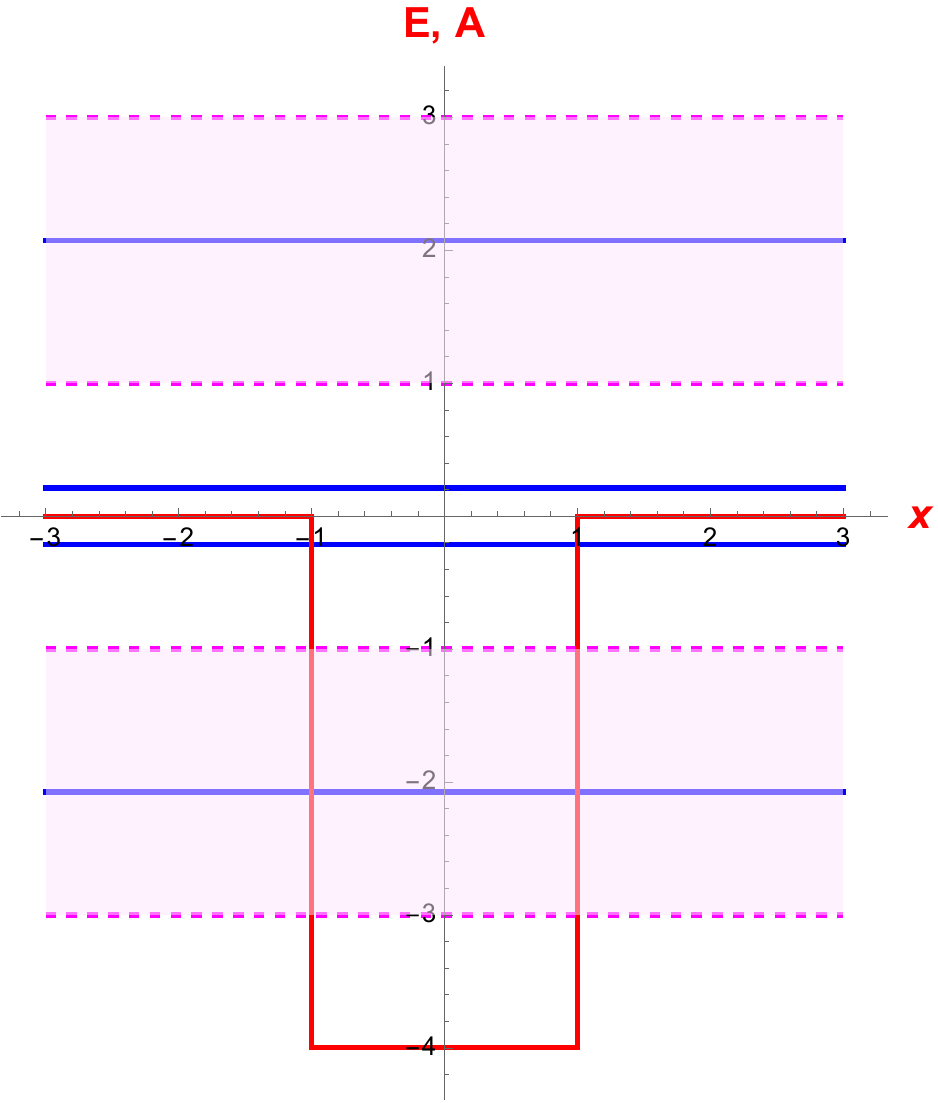}
\caption{Spectrum of the magnetic waveguide for $a_0=4$ and $k=-3$. The energy levels can be appreciated on the left and on the right graphics of Fig.~\ref{10ab}; they a determined by the cuts of vertical dotted lines. The two magenta regions between dashed lines represent the intervals of classical bound states. The ground positive and negative energies are out of these intervals and correspond to edge states.  
}
  \label{electricBxx}

\end{figure}

In Fig.~\ref{electricBxx} it is shown an example of  magnetic potential with four bound energy levels, where two of them are standard an the other two correspond to edge states. Their spectrum are symmetric with respect to the origin.  This potential is not the effective potential of an Schr\"odinger equation, so that some of the energy levels are above the potential.

\begin{figure}[h!]
  \centering
\includegraphics[width=0.5\textwidth]{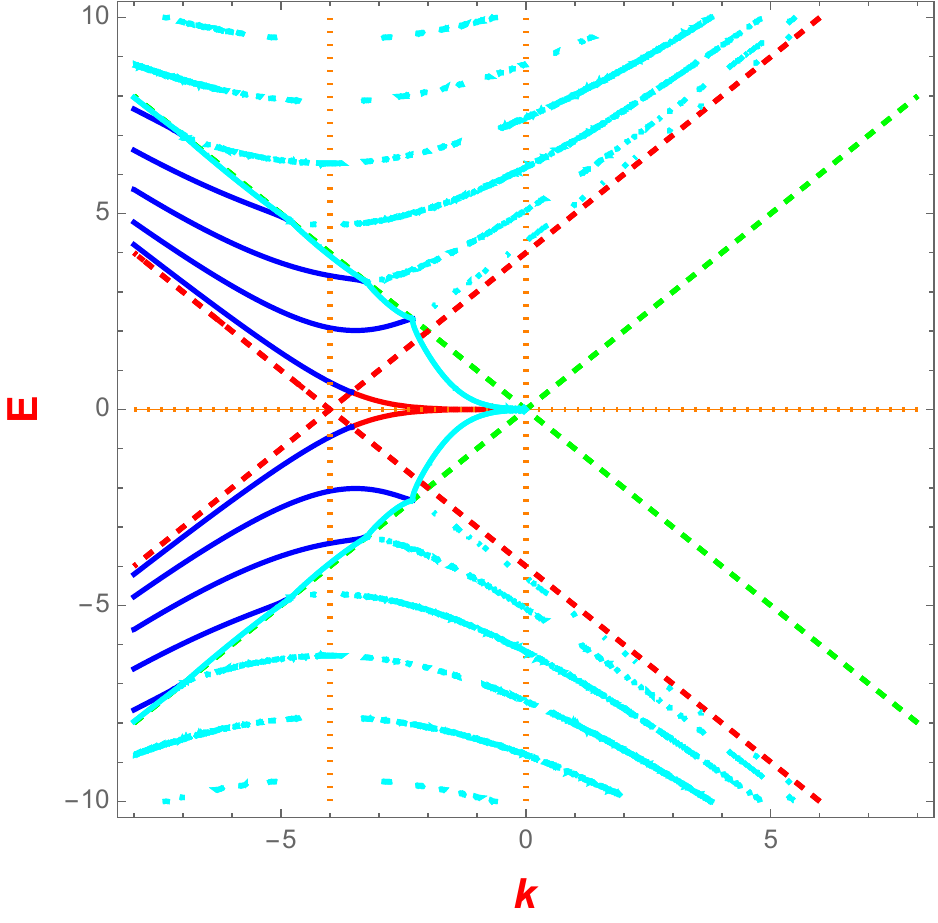}
\caption{Spectrum of the magnetic waveguide for $a_0=4$. 
The plot includes the spectral curves of bound states (blue curves),  those of resonant states $T=1$
 (cyan curves) and edge states (red curves)  for a magnetic well.  
}
  \label{electricB}
\end{figure}

Next, in figure Fig.~\ref{electricB} it is shown  two types of curves: the spectral curves of bound states together with the curves of resonant states whose transmission probability is $T=1$. These two types of curves match perfectly, so that one may interpret that the well captures resonances to turn them into bound states.

\begin{figure}[h!]
  \centering
\includegraphics[width=0.4\textwidth]{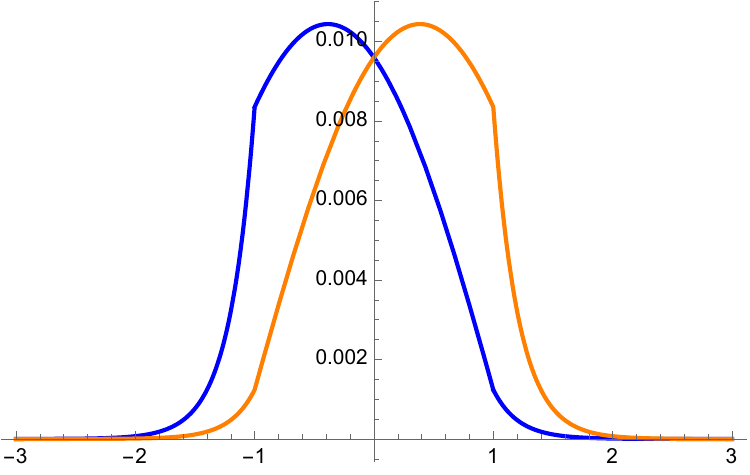}
\caption{Plot of the two components of a Dirac spinor corresponding to a standard ground state with $k=-3$, $a_0=4$. The blue one is for ${\rm Re}(\psi_1)$, the orange for ${\rm Im}(\psi_2)$.
}
  \label{11ab}
\end{figure}

\begin{figure}[h!]
  \centering
\includegraphics[width=0.4\textwidth]{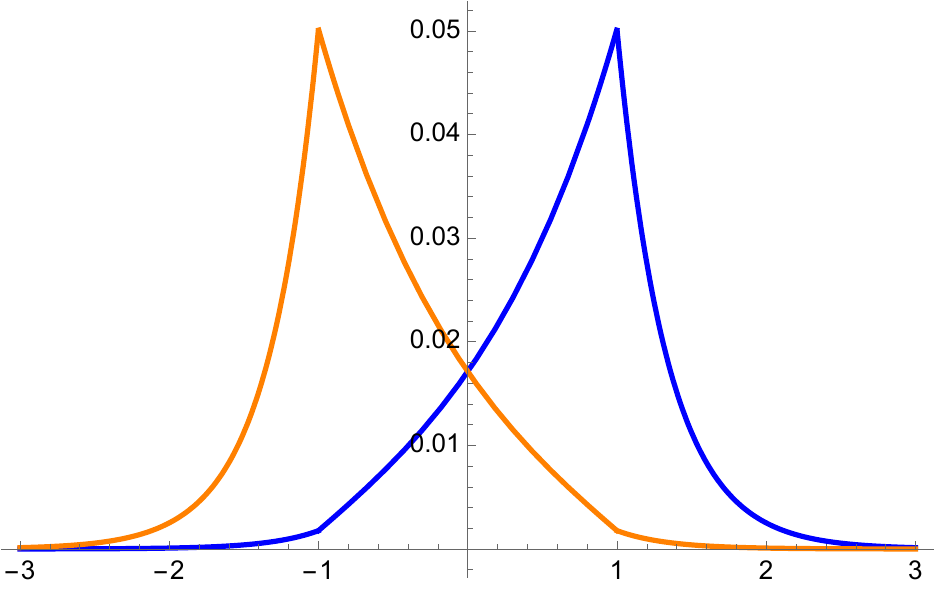}
\
\includegraphics[width=0.4\textwidth]{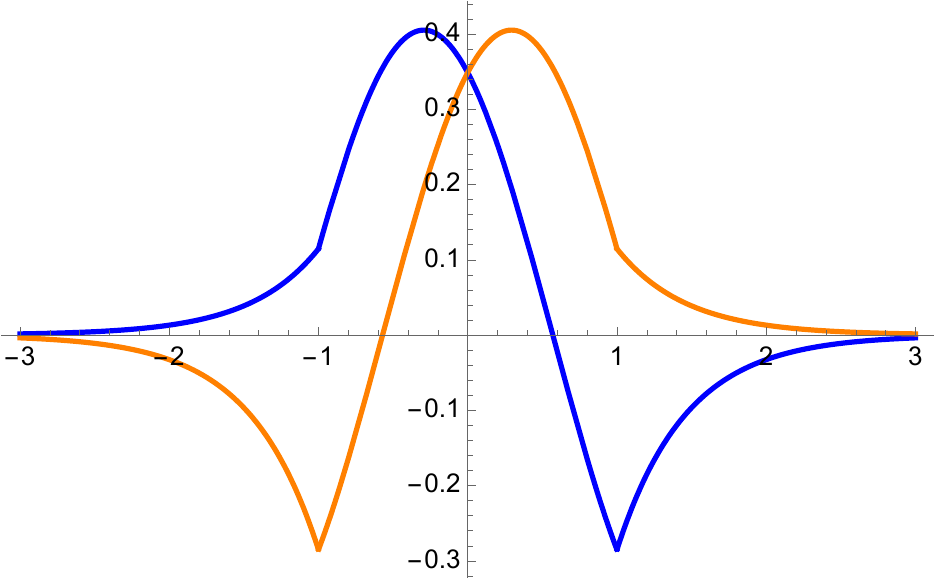}
\caption{Plots of the two components of a Dirac spinor corresponding to an edge ground state (left) and to the first (non-edge) excited state (right) for $k=-3, a_0=4$.
}
  \label{12abc}
\end{figure}

Fig.~\ref{11ab} shows the components of a standard ground state, while Fig.~\ref{12abc} (left) are the graphics of an edge ground state. We see the difference, each component of edge states  has a high peak at each of the edges of the magnetic strip. In Fig.~\ref{12abc} (right) it is shown the components of an excites states. In all the graphics Fig.~\ref{11ab}- \ref{12abc} the spinor wave functions satisfy the reflection symmetry in $x$ given in (ii) of sec.~2.1.

\subsection{Quantum electric well}
In this case we have an electric potential $V(x)$ given in (\ref{elect}), and
the Dirac--Weyl equation is
\begin{equation}\label{dirac2e}
\Big(\sigma_x(-i \partial_x)+\sigma_y(-i \partial_y)\Big)\Psi({\bf x}) = (E - V(x))\Psi({\bf x})
\end{equation}
The pseudospinor  has the form given in (\ref{psi})
where $\phi_1, \phi_2$ are constants, in (\ref{dirac2e}). We can obtain the solutions of (\ref{dirac2e}) in the three regions: region I ($x<-1$), region II ($-1<x<1$), and region III ($x>1$). Let us start with the scattering of incident states.

\subsubsection{Scattering}

In region I two independent solutions are:
\begin{itemize}
\item {\it Region I.} ($x<-1$)

In region I two independent solutions are:
\begin{equation}\label{psi1phib}
\Psi_I^{\pm}(E_\pm,{\bf x})= e^{i k y} e^{\pm i k_x x}
\left(\begin{array}{c}
1
\\[1.5ex]
\frac{(\pm k_x+ik)}{E_{\pm}}
 \end{array}\right),
 \qquad  
 k_x=\sqrt{E^2-k^2}
\end{equation}
where ${\bf x} = (x,y)$. The general solution in region I is
\begin{equation}\label{psi1phiIb}
\Psi_I(E_\pm,{\bf x})= \alpha_1\,\Psi_I^+(E_\pm,{\bf x})
+ \alpha_2\,\Psi_I^-(E_\pm,{\bf x})
\end{equation}

\item {\it Region II.} ($-1<x<1$)

We obtain $\phi_1$ and $\phi_2$ and we can write the independent solutions 
$\Psi_{II}^\pm(E_\pm,\bf x)$
\begin{equation}\label{psi2phipmb3}
\Psi_{II}^{\pm}(E_\pm,{\bf x})= e^{i k y} e^{\pm i k'_x x}
\left(\begin{array}{c}
1 
\\[1.5ex]
\frac{(\pm k'_x+i k)}{E_{\pm}}
 \end{array}\right)\,,\qquad
 k'_x=\sqrt{(E_\pm+v_0)^2-k^2}
\end{equation}
The general solution is
\begin{equation}\label{psi2phigenb}
\Psi_{II}(E_\pm,{\bf x})= \beta_1 \Psi_{II}^{+}(E_\pm,{\bf x}) +
\beta_2 \Psi_{II}^{-}(E_\pm,{\bf x})
\end{equation}
where $\beta_1,\beta_2$ are arbitrary constants.

\item {\it Region III.}  ($x>1$)

Following the same procedure as above, for this region we have the same solutions of region I:
\begin{equation}\label{psi3phi}
\Psi_{III}^{\pm}({\bf x})= e^{i k y} e^{\pm i k_x x}
\left(\begin{array}{c}
1 
\\[1.5ex]
\frac{(\pm k_x+ik)}{E_{\pm}}
 \end{array}\right)\,,
 \qquad  
 k_x=\pm\sqrt{E^2-k^2}
\end{equation}
\begin{equation}\label{psi1phiIII}
\Psi_{III}(E_\pm,{\bf x})= \gamma_1\,\Psi_{III}^+(E_\pm,{\bf x})
+ \gamma_2\,\Psi_{III}^-(E_\pm,{\bf x})
\end{equation}

where $\gamma_1,  \gamma_2$ are  the constants. The scattering state is characterized by:
$\alpha_1 = 1, \alpha_2 =r, \gamma_1= t, \gamma_0=0$, where $r$ is the reflection amplitude and $t$ is the transmission amplitude.

{\bf Remark 1.} In order to solve the scattering  problem, we must assume that $k_x$ is real   and $|E_\pm| > |k|$, to have  ingoing and outgoing plane waves in the first and third regions, respectively.

{\bf Remark 2.} Next, for second region we must take into account
 \begin{equation}\label{macht11}
{k'_x}^2=(E_\pm +v_0)^2 - k^2
\end{equation}

$\bullet$ If 
$ 
|E_\pm+ v_0|> |k|$ 

Then, the momentum $k'_x$ in region II will be real. Classically, this means that
the electron will pass the electric waveguide.  

$\bullet$ If $|E_\pm+ v_0| < |k|$ 

Then, $k'_x$ will be imaginary. This may happen, if 
 \begin{equation}\label{macht13}
 |E_\pm + v_0|< |k| < |E_\pm| 
\end{equation}
This situation can not happen for positive energies $E_+$ and positive $v_0$. So, we will not discuss it in greater detail. In the  classical context there will be total reflection. However, in the quantum case there may be transmission due to quantum tunnelling. In this case, we must take the solutions:
\begin{equation}\label{psi2phipmb2}
\Psi_{II}^{\pm}(E_\pm,{\bf x})= e^{i k y} e^{\pm k'_x x}
\left(\begin{array}{c}
1 
\\[1.5ex]
\frac{\mp i k'_x+i k}{E_{\pm}}
 \end{array}\right)
\end{equation}
For the condition (\ref{macht12}), the general solution is
\begin{equation}\label{psi2phigen}
\Psi_{II}(E_\pm,{\bf x})= \beta_1 \Psi_{II}^{+}(E_\pm,{\bf x}) +
\beta_2 \Psi_{II}^{-}(E_\pm,{\bf x})
\end{equation}
where $\beta_1,\beta_2$ are arbitrary constants.

\end{itemize}

Once obtained the general solutions in the three regions, the scattering solution
can be found in the same way as the magnetic case, taking into account the
formulas (\ref{psi1g22}), (\ref{macht12}) and (\ref{macht22}). 
After this computation with the help of {\em Mathematica}, the transmission amplitude, in terms of energy \ 
takes the form
\begin{equation}
 t(E, k, a_0) = 
\frac{ 2\exp[-2 i k_x]\, k_x k'_x}{2 k_x k'_x
       \cos(2 k'_x) - 
     i (k_x k'_x)^2 \sin(2 k'_x)} 
\end{equation}

\begin{figure}[h!]
  \centering
\includegraphics[width=0.45\textwidth]{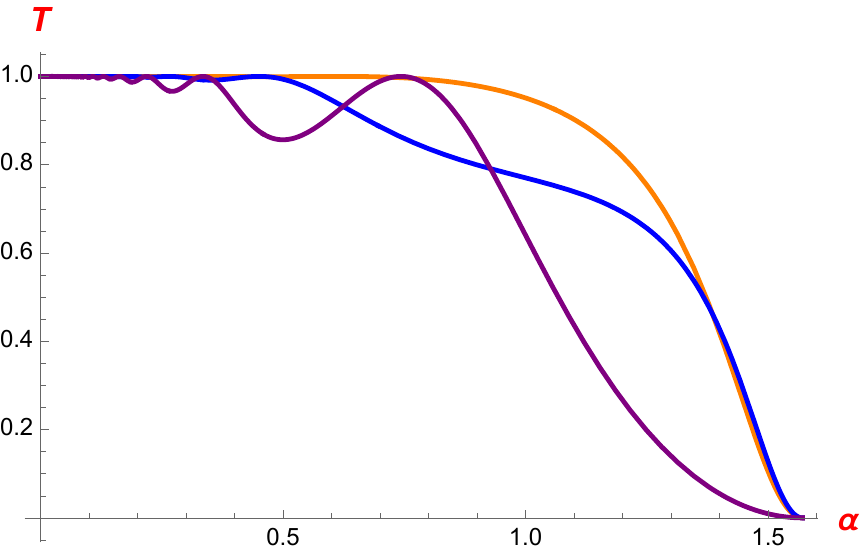}
\qquad 
\includegraphics[width=0.35\textwidth]{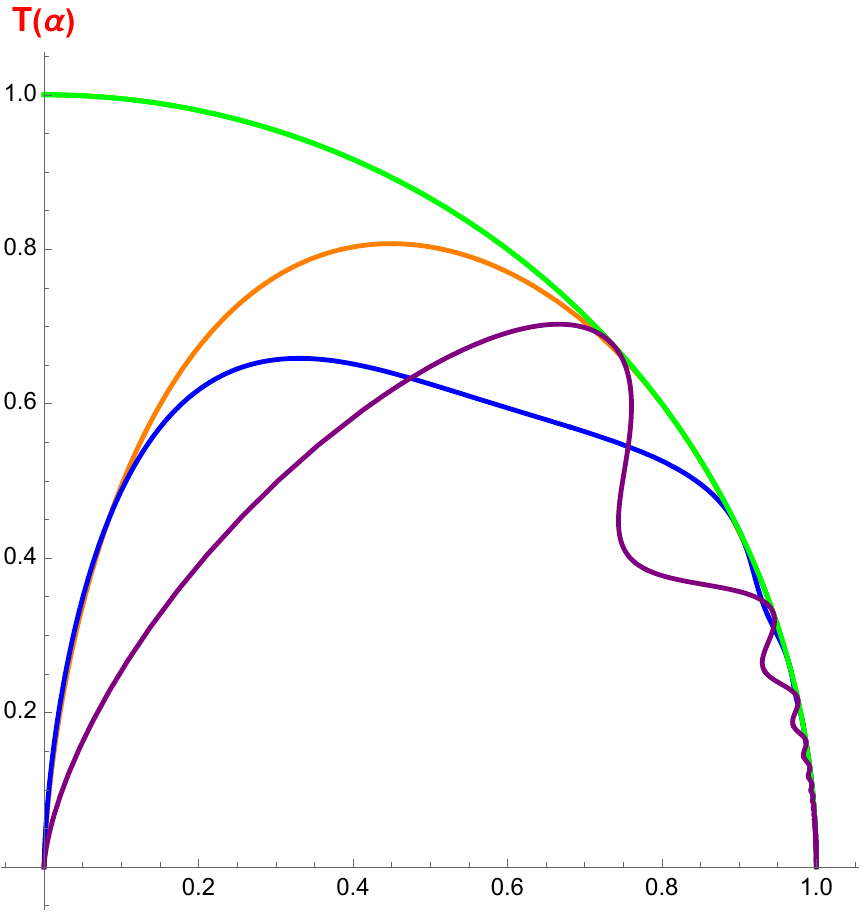}
\caption{Transmission coefficients $T(\alpha)$ of electric wells for $k=1$ as a function of $\alpha$ and for three depth values of the well: $v_0=8$ (purple), $v_0=1$ (in blue),  $v_0=1/8$ (orange).
}
  \label{figurasEs}
\end{figure}

\begin{figure}[h!]
  \centering
\includegraphics[width=0.45\textwidth]{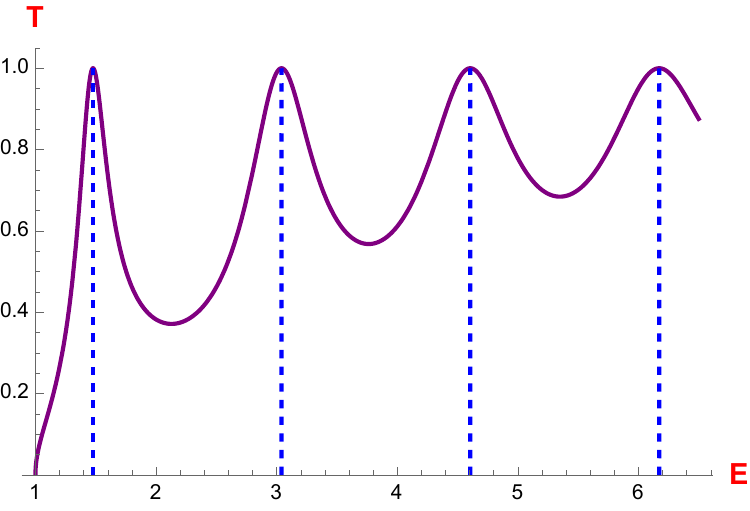}
\caption{Resonances of $T(E)$ (as a function of $E$) for $k=1$, $v_0=8$, appear for the values
$\sin(2 \sqrt{-k^2 + (E + v_0)^2})= \sin(2 k_x') = 0$.
}
  \label{figurasEs2}
\end{figure}

Some examples of the transmission coefficients for  scattering by an electric well  are given in Fig.~\ref{figurasEs} as functions of the incident angle. In this figure one can appreciate that deeper wells give rise to higher interferences and less transmission.
Fig.~\ref{figurasEs2} shows the amplitude as function of the incident energy. The resonances with $T=|t|=1$ take place for the values $\sin(2 k'_x)=0$, i.e., $2 k'_x= n\pi$,$n\in {\mathbb Z}$ \cite{peeters06}, as in the magnetic case.

\subsubsection{Bound states}

\begin{itemize}
\item {\it Region I.} ($x<-1$) 

In this region, in order to have bound states, we have the  same condition
as in the magnetic case: $k^2>E^2$ and $q_x=\pm\sqrt{k^2- E^2}$. 
Thus, we have
\begin{equation}\label{li1}
|k|>|E|
\end{equation}
Then, the solution (\ref{psi}) has the form:
\begin{equation}\label{psib}
\Psi_I({\bf x})= e^{i k y} e^{q_x x}
\left(\begin{array}{l}
\phi_1 
\\[1.5ex]
\phi_2  
 \end{array}\right)
\end{equation}
From these equations, $E^2=k^2- q_x^2$
\begin{equation}\label{psi1phib}
\Psi_I^{\pm}(E_\pm,{\bf x})= e^{i k y} e^{\pm q_x x}
\left(\begin{array}{c}
1
\\[1.5ex]
\frac{(\mp i q_x+ik)}{E_{\pm}}
 \end{array}\right),
 \qquad  
 q_x=  \sqrt{k^2-E^2}
\end{equation}
 The general solution in region I is
\begin{equation}\label{psi1phiIb}
\Psi_I(E_\pm,{\bf x})= \alpha_1\,\Psi_I^+(E_\pm,{\bf x})
+ \alpha_2\,\Psi_I^-(E_\pm,{\bf x})
\end{equation}

\item {\it Region II.} ($-1<x<1$)

Next, in the second region we have also two possibilities:
 \begin{equation}\label{macht11b}
{k'}_x^2=(E_\pm+ v_0)^2 - k^2>0,  \qquad -{q'}_x^2=(E_\pm+ v_0)^2 - k^2<0
\end{equation} 
Therefore, we get the following inequality:
\begin{equation}\label{li2}
|E_\pm+ v_0|< |k|
\end{equation}
The first possibility gives us standard bound states and the second one leads to edge states which are not present in the electric well. 

For the first possibility  we obtain $\phi_1$ and $\phi_2$ and we can write the independent solutions 
$\Psi_{II}^\pm(E_\pm,\bf x)$
\begin{equation}\label{psi2phipmb2}
\Psi_{II}^{\pm}(E_\pm,{\bf x})= e^{i k y} e^{\pm i k'_x x}
\left(\begin{array}{c}
1 
\\[1.5ex]
\frac{\pm k'_x+i k}{E_{\pm}+v_0}
 \end{array}\right)\,,\qquad
 k'_x= \sqrt{k^2-(E_\pm +v_0)^2}
\end{equation}
The general solution is
\begin{equation}\label{psi2phigenb}
\Psi_{II}(E_\pm,{\bf x})= \beta_1 \Psi_{II}^{+}(E_\pm,{\bf x}) +
\beta_2 \Psi_{II}^{-}(E_\pm,{\bf x})
\end{equation}
 For the second option of edge states:
\begin{equation}\label{psi2phipmbe}
\Psi_{II}^{\pm}(E_\pm,{\bf x})= e^{i k y} e^{\pm |q'_x| x}
\left(\begin{array}{c}
1 
\\[1.5ex]
\frac{\mp i \,q'_x+i k}{E_{\pm}+v_0}
 \end{array}\right)\,,\qquad
| q'_x|= \sqrt{k^2-(E+v_0)^2}
\end{equation}

But we will see later that in the electric case there will be no solution of this type.

\item {\it Region III.}  ($x>1$)

\begin{equation}\label{psi3phib}
\Psi_{III}^{\pm}({\bf x})= e^{i k y} e^{\pm q_x x}
\left(\begin{array}{c}
1 
\\[1.5ex]
\frac{(\mp i q_x+ik)}{E_{\pm}}
 \end{array}\right)\,,
 \qquad  
 q_x= \sqrt{k^2-e^2}
\end{equation}
\begin{equation}\label{psi1phiIIIb}
\Psi_{III}(E_\pm,{\bf x})= \gamma_1\,\Psi_{III}^+(E_\pm,{\bf x})
+ \gamma_2\,\Psi_{III}^-(E_\pm,{\bf x})
\end{equation}

\end{itemize}

\begin{figure}[h!]
  \centering
\includegraphics[width=0.4\textwidth]{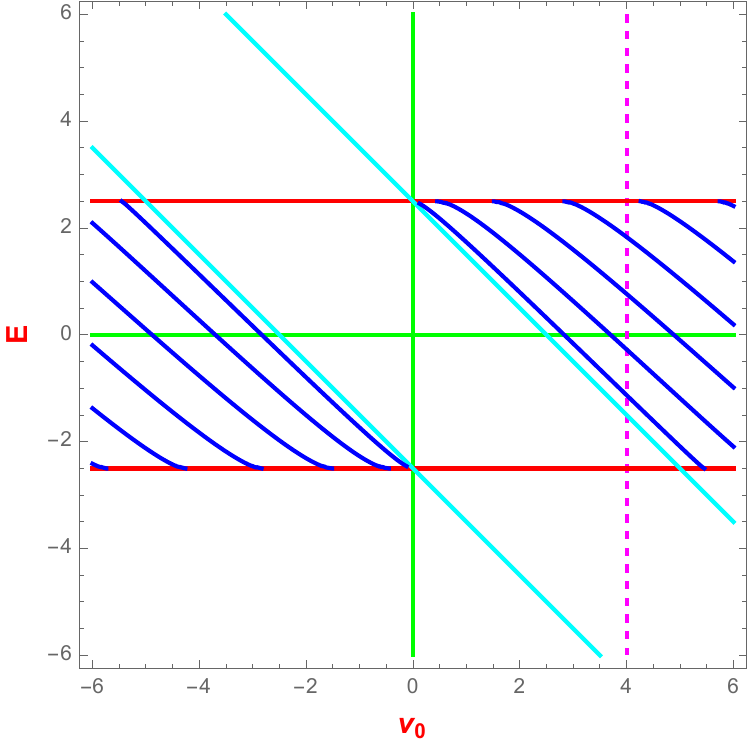}
\qquad 
\includegraphics[width=0.4\textwidth]{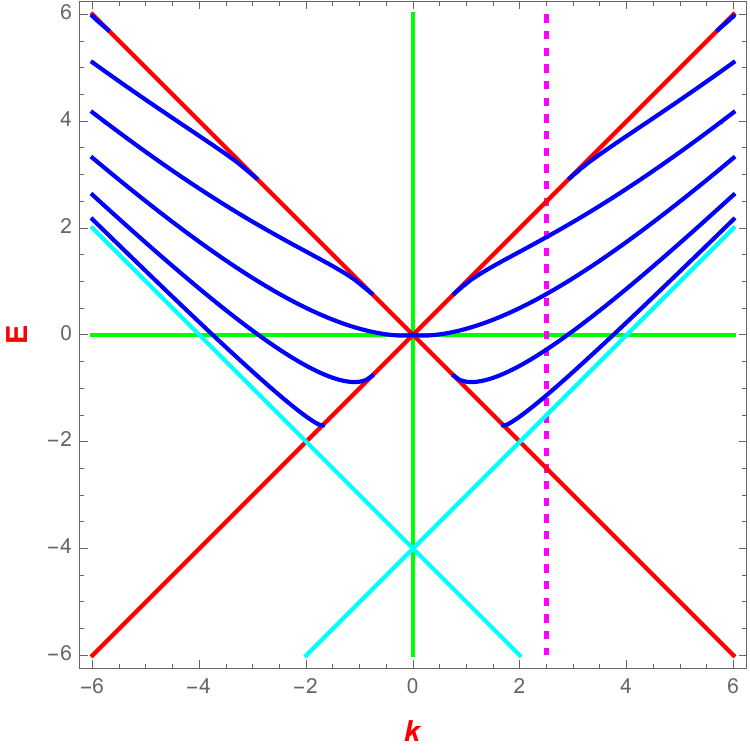}
\caption{(left) Spectral electric curves in the $E$-$v_0$ plane for $k=2.5$. (right) Spectral electric curves in the $E$-$k$ plane for $v_0=4$. Red and cyan straight lines correspond  to inequalities (\ref{li1}) and (\ref{li2}).
The vertical magenta line determines de four spectral points corresponding to the configuration $v_0=4$ and $ k=2.5$.  
Notice the $k$-symmetry (right) and $(E,v_0) \to (-E,-v_0)$ (left).}
  \label{2bounde}
\end{figure}

\begin{figure}[h!]
  \centering
\includegraphics[width=0.4\textwidth]{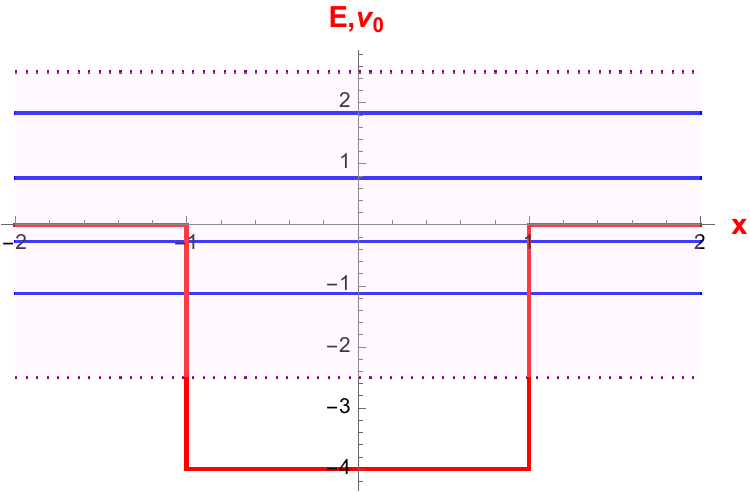}
\caption{Spectrum of the electric waveguide for $v_0=4$ and $k=2.5$ corresponding to the magenta line of Fig.~\ref{2bounde}.  
}
  \label{electricBb}
\end{figure}

\begin{figure}[h!]
  \centering
\includegraphics[width=0.4\textwidth]{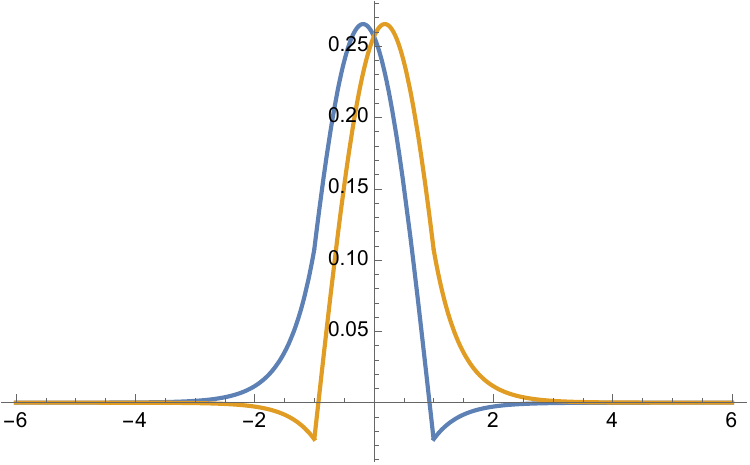}
\qquad
\includegraphics[width=0.4\textwidth]{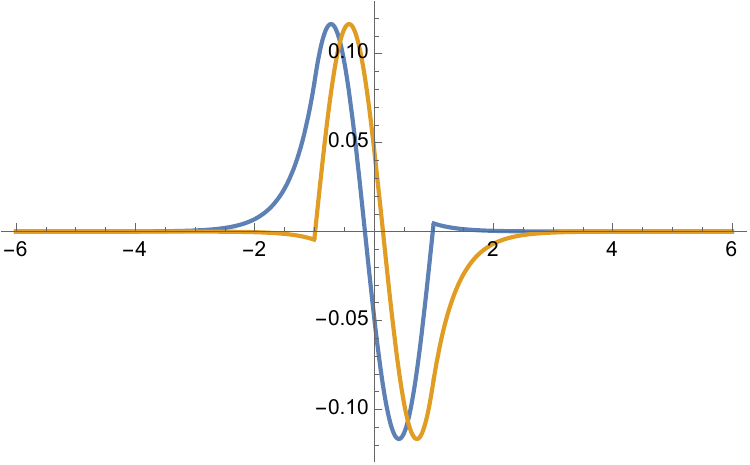}
\caption{Plot of the two components of a Dirac spinor corresponding to the two lower (negative) levels of $k=2.5$, $v_0=4$: $E_0=-1.12$ (left), $E_1=-0.27$  (right).
}
  \label{electricB01}
\end{figure}

\begin{figure}[h!]
  \centering
\includegraphics[width=0.4\textwidth]{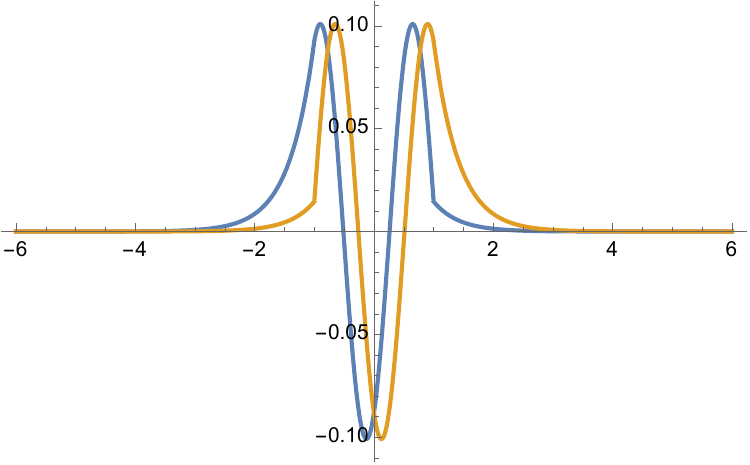}
\qquad
\includegraphics[width=0.4\textwidth]{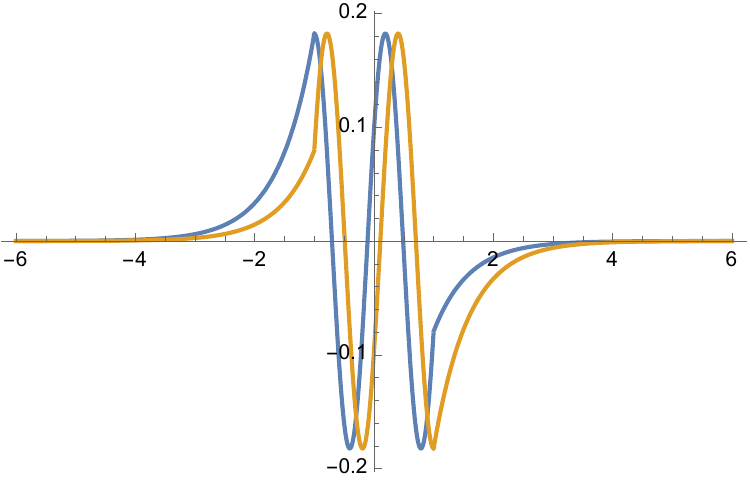}

\caption{Plot of the two components of a Dirac spinor corresponding to the two upper (positive) levels for $k=2.5$, $v_0=4$: $E_2=0.76$ (left), $E_3=1.83$ (right).
}
  \label{electricB23}
\end{figure}

Fig.~\ref{2bounde} includes the spectral curves in the plane energy-depth ($v_0$) of the well (left) and corresponding to the plane energy-vertical momentum, $k$ (right). We have selected the spectrum for a particular configuration: $k=2.5, v_0=4$. We have computed the energy values (see Fig.~\ref{electricBb}, where it is represented the electric well together with the energy levels) and plotted the corresponding spinor eigenfunctions in Figs.~\ref{electricB01}-\ref{electricB23}. All of them have $x$-reflection symmetry. It happens that the two lowest eigenvalues are negative and their corresponding eigenfunctions have a partial edge character (a cusp at the border of the well). The two upper eigenvalues are positive and they have the standard form. It is clear the difference with the true edge state of Fig.~\ref{12abc}, for the magnetic waveguide.

\begin{figure}[h!]
  \centering
\includegraphics[width=0.33\textwidth]{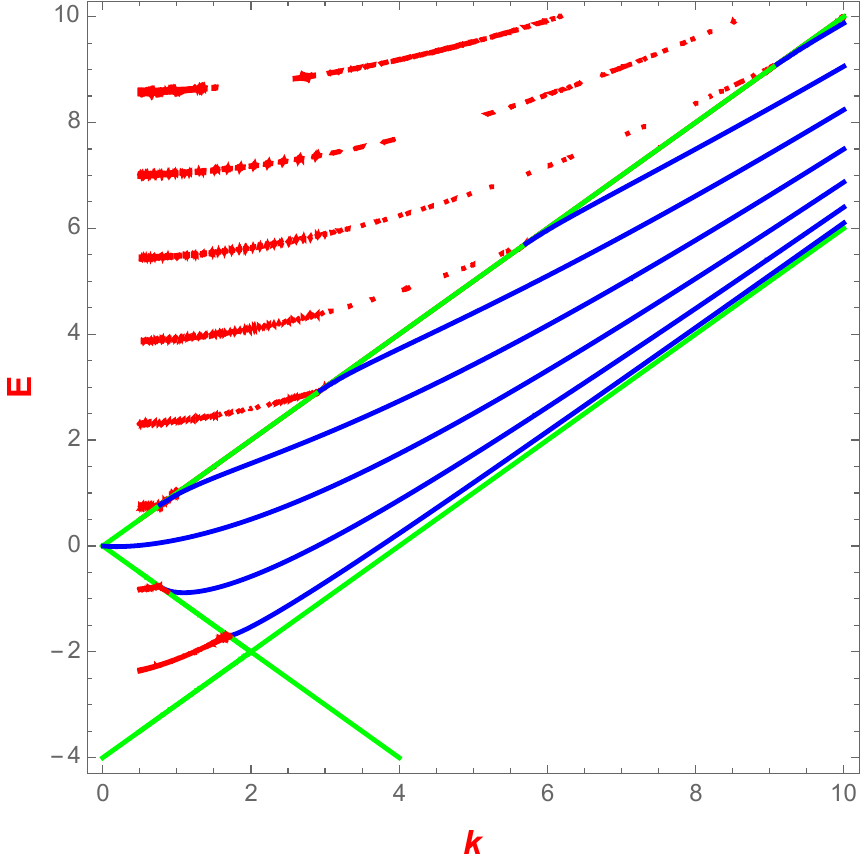}
\ 
\includegraphics[width=0.4\textwidth]{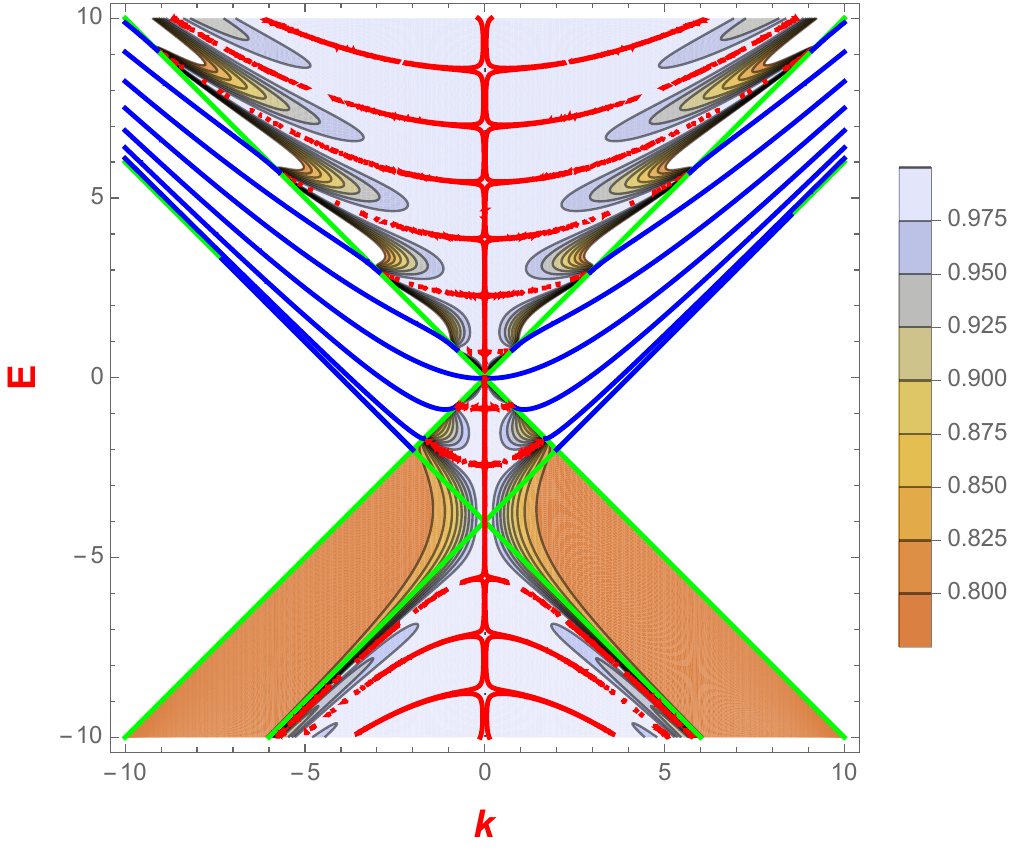}
\caption{(left) Plot of a detail of spectral curves of bound states (in blue) and resonant states (in red) for an electric well, for 
$v_0= 4$. (right) Complete picture of spectral curves (in blue), resonant curves (in red) and global values of transmission coefficient for this electric well. 
}
  \label{2bounde2}
\end{figure}

Finally, Fig.~\ref{2bounde2} shows the spectral curves of bound states and the curves of the resonant scattering states which have the highest transmission coefficient $T=1$ (see also \cite{peeters06}). These two kinds of curves match continuously where the resonant states turn into bound states.
This is a representation of the phenomenon  of capture and collapse processes which is produced for electrons and holes.

\section{Conclusions}

In this work,  we have dealt with Dirac particles
(relativistic massless, spin-1/2) of planar Dirac materials like graphene, 
under external electric and magnetic waveguide potentials in a quite comprehensive way including a classical relativistic model as introduction.
We have paid attention to the symmetries that are the main ingredient of our program. They are made explicit through the graphics and are responsible of qualitative differences. 
This presentation contains  complementary comparisons which allow for a better understanding of electric and magnetic waveguides: i) Classical approximation to waveguide systems. ii) Bounded and scattering states in quantum waveguides. iii) The superposition of classical and quantum properties on scattering and bound states. 

All the results, as we mentioned along this work were obtained 
by explicit computations and also represented a series of appropriate  graphics. Thus, we hope that the main features of these waveguides can be easily appreciated from these figures. In particular,  we want to remark the importance of those corresponding to the energy-momentum ($E$-$k$) and energy-intensity ($E$-$v_0/a_0$) which are complementary and exhibit the different symmetries of electric and magnetic systems. In the near future, we plan to complet this program for waveguides of simultaneous electric and magnetic potential interactions.


\section*{Acknowledgments}

We appreciate the support of the QCAYLE project, funded by the European Union--NextGenerationEU, and PID2020-113406GB-I0 project funded by the MCIN of Spain.
\c{S}.~K. thanks Ankara University and the warm hospitality of the Department of Theoretical Physics of the University of Valladolid, where part of this work has been carried out, and to the support of its GIR of Mathematical Physics.

\end{document}